\documentclass{jfm}
\usepackage{graphicx}
\usepackage{epstopdf, epsfig}
\usepackage{newtxtext}
\usepackage{newtxmath}
\usepackage{natbib}

\shorttitle{Charged inertial particles in turbulent channel flow}
\shortauthor{X. Ruan, M. X. Diaz-Lopez, M. Gorman and R. Ni}

\title{Direct numerical simulations on transport and deposition of charged inertial particles in turbulent channel flow}

\author{Xuan Ruan,
  Miguel X. Diaz-Lopez,
  Matthew T. Gorman
 \and Rui Ni \corresp{\email{rui.ni@jhu.edu}}}

\affiliation{Department of Mechanical Engineering, Johns Hopkins University, Baltimore, MD 21218, USA}

\begin{document}

\maketitle

\begin{abstract}
From particle lifting in atmospheric boundary layers to dust ingestion in jet engines, the transport and deposition of inertial particles in wall-bounded turbulent flows are prevalent in both nature and industry. Due to triboelectrification during collisions, solid particles often acquire significant charges. However, the impacts of the resulting electrostatic interaction on particle dynamics remain less understood. In this study, we present four-way coupled simulations to investigate the deposition of charged particles onto a grounded metal substrate through a fully developed turbulent boundary layer. Our numerical method tracks the dynamics of individual particles under the influence of turbulence, electrostatic forces, and collisions. We first report a more pronounced near-wall accumulation and an increased wall-normal particle velocity due to particle charging. In addition, contrary to predictions from the classic Eulerian model, the wall-normal transport rate of inertial particles is significantly enhanced by electrostatic forces. A statistical approach is then applied to quantify the contributions from turbophoresis, biased sampling, and electrostatic forces. For charged particles, a sharper gradient in wall-normal particle fluctuation velocity is observed, which substantially enhances turbophoresis and serves as the primary driving force of near-wall particle accumulation. Furthermore, charged particles are found to sample upward-moving fluids less frequently than neutral particles, thereby weakening the biased sampling effect that typically pushes particles away from the wall. Finally, the wall-normal electric field is shown to depend on the competition between particle-wall and particle-particle electrostatic interactions, which helps to identify the dominant electrostatic force across a wide range of scenarios.
\end{abstract}


\section{Introduction}
\label{sec:headings}
The transport of charged inertial particles in wall-bounded turbulent flows occurs across a wide range of natural and industrial processes. Common examples include electrified dust storms \citep{ZhengJGR2004, ZhangNatComm2020}, gas-solid fluidized beds \citep{PeiPT2016}, dust ingestion in jet engines \citep{ShinozakiAEM2013, Diaz-LopezAIAA2025}, and powder delivery systems \citep{GrosshansPT2016}. In these processes, solid particles easily accumulate electrical charges through frequent particle-particle or particle-wall collisions \citep{GrosshansJFM2017, LacksNRC2019}. The resulting electrostatic forces could drastically influence particle dynamics, including enhancing dust emission in atmospheric boundary layers \citep{KokPRL2008, EspositoGRL2016}, accelerating particle transport in pipe flows \citep{GuhaARFM2008, YaoPT2021}, initiating particle aggregation and deposition growth \citep{LeeNatPhys2015, SippolaJFM2018, RuanJCP2022, GormanPRE2024}, and inducing turbulent modulations \citep{CuiJFM2024}. Moreover, the electric field generated by tribocharged particles may exceed the breakdown limit and trigger electrical discharges, posing potential risks to equipment and personnel safety \citep{EckhoffBook2003, DiRenzoNC2018}. Therefore, investigating the dynamics of charged particles is crucial for revealing the role of electrostatic interactions and advancing our knowledge of the widespread electrostatic phenomena in particle-laden flows.

The transport of neutral inertial particles in wall-bounded flows has been extensively studied and essential physical processes have been revealed \citep{SoldatiIJMF2009, BrandtARFM2022}. The presence of a wall creates a significant gradient of turbulence intensity in the wall-normal direction, driving inertial particles to preferentially migrate towards the wall, which is known as the turbophoresis effect \citep{CaporaloniJAS1975, ReeksJAS1983}. Both numerical and experimental studies have shown that the near-wall particle transport is dominated by buffer-layer coherent structures \citep{NintoJFM1996, MarchioliJFM2002}. In particular, quasi-streamwise vortices generate sweeps and ejections. Inertial particles brought towards the wall by sweeps are trapped in the viscous layer until they are re-entrained into the outer layer by ejections. As ejection-induced re-entrainment is less efficient, inertial particles tend to accumulate near the wall leading to the high local concentration. Moreover, the response of inertial particles to the near-wall coherent structures depends on the viscous Stokes number $St^+$, which is defined as the ratio of the particle relaxation time to the viscous time scale, and the strongest near-wall particle accumulation is observed for $St^+=10-50$ \citep{SardinaJFM2012}. After reaching equilibrium, particles oversample fluid motions departing from the wall to balance the turbophoresis drift towards the wall \citep{PicciottoPOF2005, PicanoPOF2009, JohnsonJFM2020}. In addition, near-wall particles are also found to form elongated streaky structures, corresponding to the low-speed fluid streaks accompanying quasi-streamwise vortices \citep{RousonJFM2001}. The dimension of such particle streaks goes up to 500-1000 wall units in the streamwise direction and are spaced by around 100 wall units in the spanwise direction \citep{SardinaJFM2012, FongJFM2019}. With the increase of the Reynolds number, the scale separation between the small-scale and large-scale structures becomes more significant \citep{HutchinsJFM2007}, and large-scale structures located in the outer layer are expected to also contribute to particle transport and accumulation. As a result, while the dynamics of particles with an intermediate $St^+$ still correlate with the near-wall vortices, particles with much larger inertia are predominantly driven by large-scale quasi-streamwise vortices whose timescale is comparable to the particle relaxation time, resulting in the formation of multiscale particle streaks in high-Reynolds-number wall-bounded turbulence \citep{WangJFM2019, BerkJFM2020, JieJFM2022, MotooriJFM2022, BerkJFM2023}.

\begin{figure}
    \centering
    \includegraphics[width=10cm]{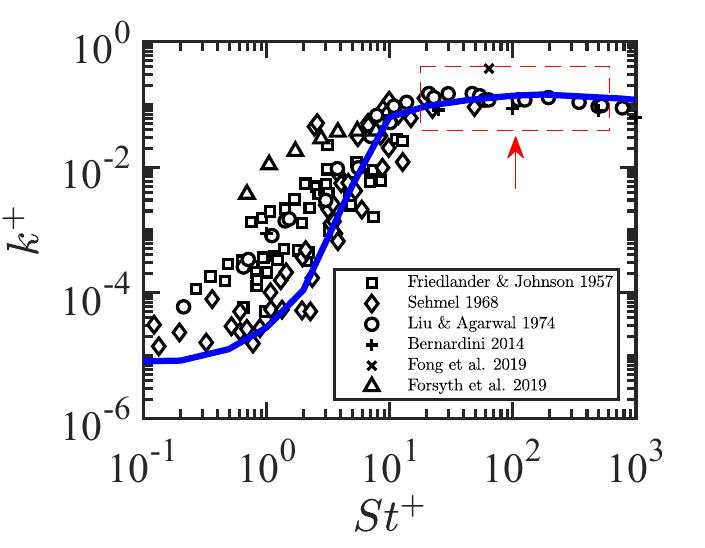}
    \caption{Dimensionless deposition velocity $k^+$ for neutral particles as a function of the particle Stokes number $St^+$ in previous works. Experimental data are plotted as scatters ($\square$: \citet{FriedlanderIEC1957}, $\diamond$: \citet{SehmelAOH1968}, $\circ$: \citet{LiuJAS1974}, $+$: \citet{BernadiniJFM2014}, $\times$: \citet{FongJFM2019}, $\triangle$: \citet{ForsythJT2019}), while the model prediction by \citet{GuhaARFM2008} is shown as the blue solid line.}
    \label{fig:DepVel_Refs}
\end{figure}

As a result of the complex particle-turbulence interaction, the particle deposition velocity at the walls, which is the primary focus of this study, varies significantly with changes in particle inertia. Figure \ref{fig:DepVel_Refs} presents the dimensionless deposition velocity from previous experimental data for neutral particles, along with the prediction based on the model of \citet{GuhaARFM2008} represented by the blue solid line. The dimensionless deposition velocity $k^+=k/u_{\tau}C_0$ is defined as the flux of particles deposited onto the wall, $k$, normalized by the average particle concentration $C_0$ and the friction velocity $u_{\tau}$. $St^+$ is the particle Stokes number defined based on the viscous scales. The experimental data exhibits considerable scatter, spanning several orders of magnitude, which was hypothesized to result from differences in particle charges across experiments. Furthermore, data points within the inertial-particle regime (highlighted by the red window in figure \ref{fig:DepVel_Refs}) are sparse. However, particles within this regime are highly relevant to problems such as dust ingestion and sandstorms, which will be further investigated in this study.

Once particles are charged, the resulting electrostatic interaction makes inertial particle behavior more complex. Most existing studies on the dynamics of charged particles in turbulence are conducted in homogeneous isotropic turbulence (HIT). In HIT, the absence of walls means that particle charging only results in the particle-particle (PP) Coulomb force. Under this condition, the significance of the Coulomb force has been quantified using both velocity- and energy-based dimensionless parameters in previous studies. The velocity-based parameter is determined by comparing the electrical migration velocity to the turbulent drift velocity \citep{LuNJP2010, LuPOF2015, DiRenzoNC2018}, while the energy-based parameter compares the electric potential energy to the particle kinetic energy \citep{LuPRL2010, BoutsikakisJCP2023, RuanJFM2024}. When electrostatic effects dominate, both the clustering and relative motion of charged particles are significantly altered \citep{KarnikPOF2012, YaoPRF2018, RuanJFM2021, BoutsikakisJFM2022}.

In wall-bounded domains, the electrostatic effects become more complicated because, in addition to the PP electrostatic interaction mentioned above, the particle-wall (PW) electrostatic interaction also plays a role. In \citet{GuhaARFM2008}, the Eulerian model is extended to account for charged particles under two key assumptions: (1) the particle velocity is modulated solely by the image force, and (2) the particle concentration remains unchanged. Using the image charge model, a charged particle near a conducting wall is subject to the Coulomb force from its own image with the opposing charge at the symmetric location about the wall. The PW interaction is thus attractive, pushing particles towards the wall and increasing particle deposition velocity. However, the electrostatic force is only found to enhance particle deposition for weak-inertia particles with $St^+ \leq 10$, while the deposition of moderate- and large-inertia particles is almost unaffected. The electrostatic-enhanced deposition of small-inertia particles is also confirmed by later direct numerical simulations, where a comprehensive numerical framework is proposed to calculate both PP and PW interactions acting on each particle \citep{YaoPT2021}. Meanwhile, when studying the wall-normal accumulation of identically charged particles, \citet{DiRenzoPRF2019} suggests that it is the collective self-induced electric force (i.e., the PP repulsion) that drives particles towards the wall. And in the later work by \citet{ZhangJFM2023} that studies the behavior of bidispersed oppositely charged particles, the PP attraction between different particle groups was found to be essential in determining the wall-normal particle distribution compared to the monodispersed case.

Despite these recent advances, several questions still remain unresolved. First, while the classic Eulerian framework by \citet{GuhaARFM2008} suggests that electrostatic forces only enhance the deposition of small-inertia particles, recent findings by \citet{ZhangJFM2023} indicate that the dynamics of large-inertia particles could also be significantly affected. This raises the question of whether, and how, electrostatic forces promote the transport and deposition of large-inertia particles. In addition, although both particle-wall (PW) and particle-particle (PP) electrostatic interactions have been found to drive charged particles towards the wall, the relative importance of these two interactions under different conditions has not been thoroughly discussed, and it remains unclear how they each contribute to the overall electrostatic force acting on charged particles. Hence, both assumptions that \citet{GuhaARFM2008} has adopted to account for the influence of the electrostatic forces require further examination.

To address the above questions, we perform four-way coupled simulations in this study. The paper is structured as follows. The simulation conditions and the numerical methods are described in section \ref{sec:Methods}. In section \ref{sec:Deposition}, we first present the effects of electrostatic forces on the wall-normal distribution and the mean velocity of charged particles, followed by a discussion on the wall-normal particle deposition velocity. A statistical approach is introduced in section \ref{sec:Model} to quantify the contributions of turbophoresis, biased sampling and electrostatic forces to wall-normal particle distribution. Section \ref{sec:IndirectEffect} then provides a detailed explanation of how turbophoresis and biased sampling are modulated. Finally, the competition between PW and PP electrostatic interactions in determining the wall-normal electric field is elucidated in section \ref{sec:Efield}.

\section{Numerical methods}
\label{sec:Methods}

\subsection{Particle parameters}
\label{sec:ParParam}
Appropriate parameters for solid particles should be selected to ensure that the particle dynamics fall within the regime relevant to real applications. The aerodynamic response of solid particles to wall-bounded turbulent flows is usually characterized by the viscous Stokes number, defined as the ratio of the particle relaxation time $\tau_p (=\rho_p d_p^2/18 \rho_f \nu_f)$ to the viscous timescale $\tau_\nu$

\begin{equation}
    St^+ = \frac{\tau_p}{\tau_\nu}=\frac{\rho_p}{18 \rho_f} \left( \frac{d_p}{\nu_f /u_\tau} \right)^2
\end{equation}

\noindent
Here, $\rho_p$ and $d_p$ are the particle density and diameter, $\rho_f$ and $\nu_f$ are the fluid density and kinematic viscosity. $u_\tau$ denotes the friction velocity.

\begin{figure}
    \centering
    \includegraphics[width=6.5cm]{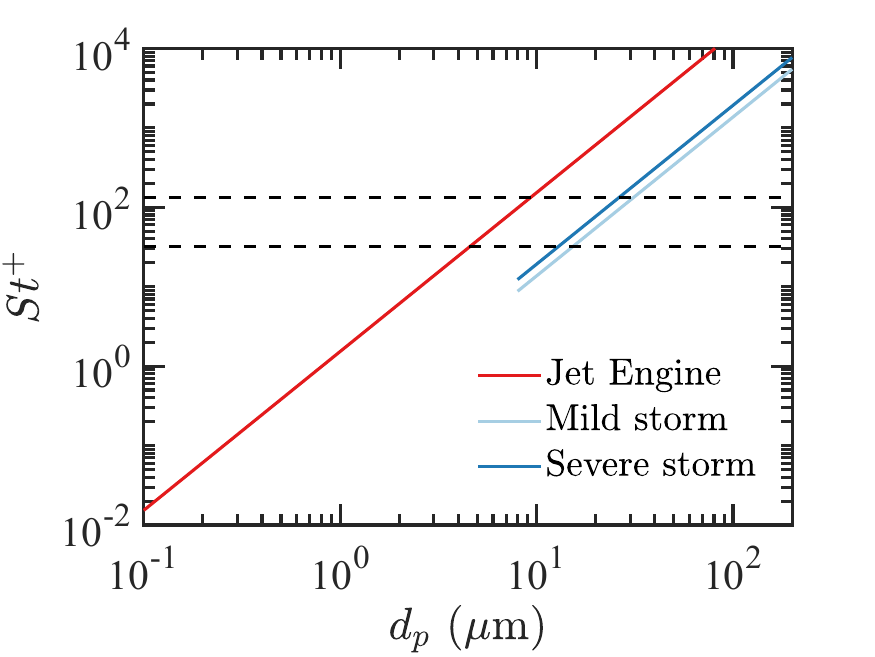}
    \caption{Dependence of particle Stokes number $St^+$ on particle size $d_p$ in different applications. Horizontal dashed lines denote $St^+=32$ and $St^+=133$.}
    \label{fig:St}
\end{figure}

For the deposition of ash particles in jet engines, typical parameters are chosen based on previous works \citep{TaylorJT1990,LawsonJT2011, LawsonJT2012, ShinozakiAEM2013, SaccoIEGTP2018} and are listed in table \ref{tab:Dust ingestion}. The friction factor $f=0.012$ is determined by the Reynolds number $Re=\rho_f U b/ \mu_f$ and the relative roughness $\epsilon_s/D_h$, where the hydraulic diameter $D_h$ is assumed to be comparable to the chord length $b$. The friction velocity can thus be estimated as $u_\tau=U \sqrt{f/8}=3.59 \ \mathrm{m/s}$. Using the ash particle density $\rho_p=1980 \ \mathrm{kg/m^3}$ and the ash particle diameter $d_p=0.1-100 \ \mathrm{\mu m}$ results in a Stokes number range of $St^+=10^{-2}-10^4$ (red line in figure \ref{fig:St}).

\begin{table}
  \begin{center}
\def~{\hphantom{0}}
  \begin{tabular}{lll}
    Parameters & Values & Units\\
    \\
    Gas velocity, $U$ & $93$ & $\mathrm{m/s}$\\
    Gas temperature, $T$ & $1500$ & $\mathrm{K}$\\
    Gas pressure, $p$ & $14$ & $\mathrm{bar}$\\
    Gas dynamic viscosity, $\mu_f$ & $5.55 \times 10^{-5}$ & $\mathrm{Pa \cdot s}$\\
    Chord length, $b$ & $0.218$ & $\mathrm{m}$ \\
    Surface roughness on blades, $\epsilon_s$ & $6 \times 10^{-6}$ & $\mathrm{m}$\\
    Reynolds number, $Re$ & $1.2 \times 10^6$ & $-$\\
    Friction factor, $f$ & $0.012$ & $-$\\
    Friction velocity, $u_\tau$ & $3.59$ & $\mathrm{m/s}$\\
    Ash particle density, $\rho_p$ & $1980$ & $\mathrm{kg/m^3}$\\
    Ash particle diameter, $d_p$ & $0.1-100$ & $\mathrm{\mu m}$\\
  \end{tabular}
  \caption{Parameters for dust ingestion problem.}
  \label{tab:Dust ingestion}
  \end{center}
\end{table}

Additionally, for the transport of dust particles in atmospheric boundary layers, the particle Stokes number can be estimated from field measurement data by \citet{ZhangNatComm2023}. The friction velocity is $u_\tau=0.54 \ \mathrm{m/s}$ for a mild sandstorm and $u_\tau=0.64 \ \mathrm{m/s}$ for a severe one. Most dust particles lie within the size range $8-200 \ \mathrm{\mu m}$. Assuming a typical dust particle density of $\rho_p=2500 \ \mathrm{kg/m^3}$, $St^+$ ranges from $O(10^1)$ to $O(10^4)$ (blue lines in figure \ref{fig:St}).

Consequently, we choose two typical Stokes numbers, $St^+=32$ and $St^+=133$ (dashed lines in figure \ref{fig:St}), which are relevant to both applications. Here, moderate-inertia particles with $St^+=32$ are more responsive to near-wall coherent structures, while large-inertia particles with $St^+=133$ exhibit more ballistic behavior \citep{JieJFM2022}.

Furthermore, the surface charging density of tribocharged particles is approximately $\sigma_c \sim 10^{-5} \ \mathrm{C/m^2}$ \citep{LeeNatPhys2015}. For typical dust particles with sizes in the tens of microns, the particle charge is around $10^{-15}-10^{-14} \ \mathrm{C}$. As a result, the particle charge $q$ in the simulations is set around this level, which is comparable to values used in previous studies \citep{ZhangJFM2023, RuanJFM2024}. In addition, since our focus is on the effects of electrostatic force, other significant forces, such as gravity and lift force \citep{MarchioliIJMF2007, BerkJFM2020, GaoJFM2024}, are not included in this study.

\subsection{Simulation system}
\label{sec:system}
\begin{figure}
    \centering
    \includegraphics[width=10cm]{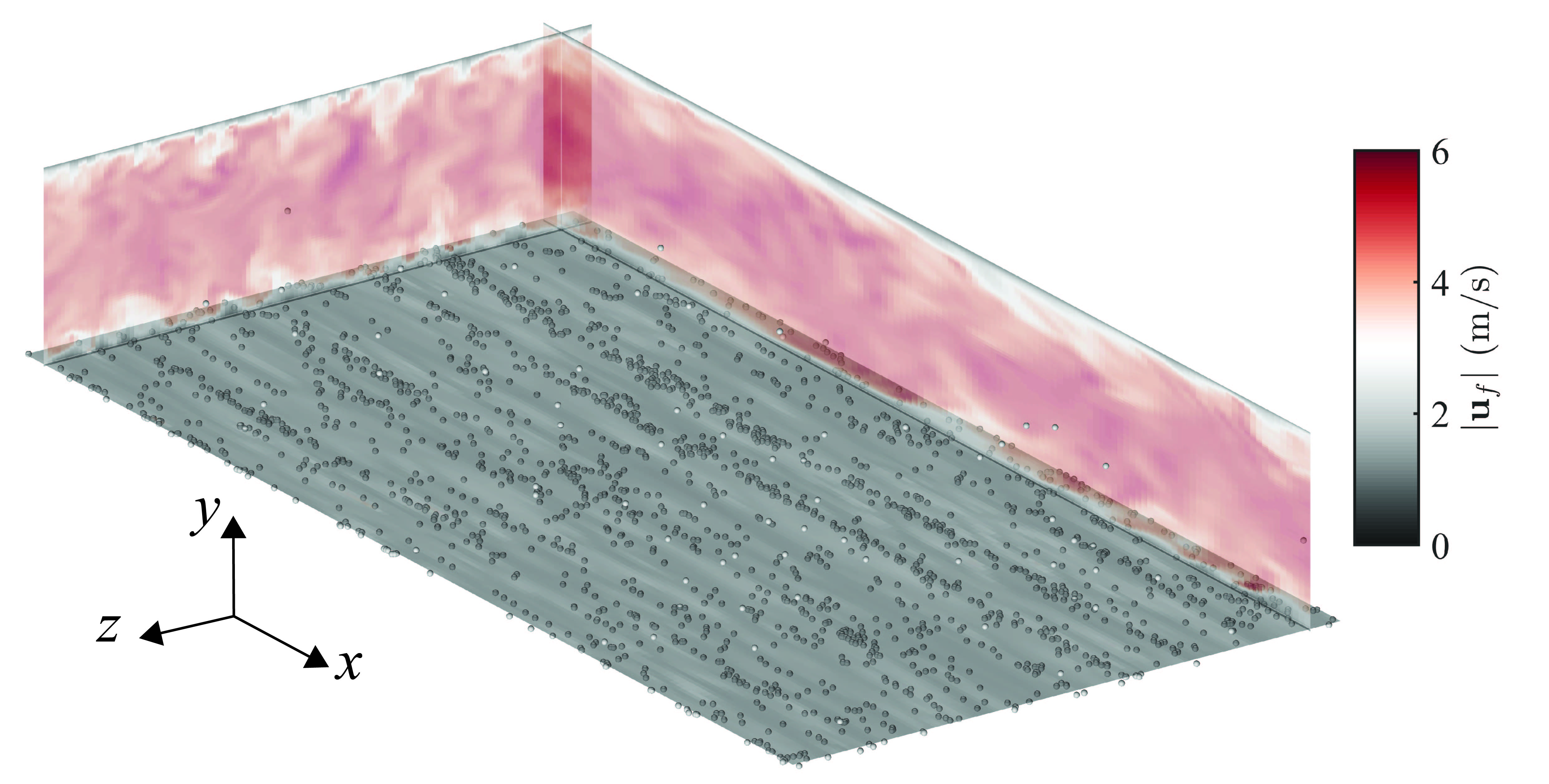}
    \caption{Snapshot of the simulation system. The color bar represents the magnitude of the fluid velocity $|\mathbf{u}_f|$. Particles are plotted as gray spheres with exaggerated sizes. For clarity, only a small portion of particles near the bottom wall are shown.}
    \label{fig:System}
\end{figure}

As shown in Fig. \ref{fig:System}, the simulation system is a particle-laden turbulent channel flow between two infinite parallel walls, and the simulation parameters are listed in Table \ref{tab:SimulationParameters}. The dimension of the computation domain is $L_x \times L_y \times L_z = 4\pi \delta \times 2 \delta \times 2\pi \delta$ with $\delta=0.01 \ \mathrm{m}$ being the half channel height. The periodic boundary condition is applied to both the streamwise ($x$) and spanwise ($z$) directions, while the no-slip boundary condition is applied to the wall-normal direction ($y$). The constant bulk velocity of the fluid phase is $U_b=4.2 \ \mathrm{m/s}$, and the friction Reynolds number is $Re_{\tau}=u_{\tau} \delta/\nu_f =180 $ with $u_{\tau}$ and $\nu_f$ being the friction velocity and the fluid kinematic viscosity, respectively. The grid number is $N_x \times N_y \times N_z = 128^3$. The grid is uniform in both x and z directions, and the nonuniform wall-norm grid is defined by the hyperbolic tangent function with the stretching factor $S=1.9$ \citep{MarchioliIJMF2008}. This leads to a grid spacing of $\Delta x^+ = 17.67$, $\Delta z^+ = 8.84$, and $\Delta y^+ = 0.49-5.58$. The grid resolution has been assessed in Appendix \ref{sec:Appd_GridResolution}, and is shown to be sufficient for the fluid flow investigated in this study. Hereinafter, variables normalized by the wall units (i.e., the friction velocity $u_{\tau}$, the viscous length scales $\delta_{\nu}=\nu_f / u_{\tau}$, and the viscous time scale $\tau_{\nu}=\nu_f / u_{\tau}^2$ ) are presented with the superscript $+$. 

\begin{table}
  \begin{center}
\def~{\hphantom{0}}
  \begin{tabular}{lll}
    Parameters & Values & Units \\
    \\
    \textit{Fluid phase} \\
    Fluid density, $\rho_f$ & $1.2$ & $\mathrm{kg/m^3}$ \\
    Fluid kinetic viscosity, $\nu_f$ & $1.5 \times 10^{-5}$ & $\mathrm{m^2/s}$ \\
    Bulk velocity, $U_b$ & $4.2$ & $\mathrm{m/s}$ \\
    Friction velocity, $u_\tau$ & $0.27$ & $\mathrm{m/s}$ \\
    Friction Reynolds number, $\mathrm{Re}_{\tau}$ & $180$ & $-$ \\
    \\
    \textit{Particle phase} \\
    Particle diameter, $d_p$ & $20$ & $\mathrm{\mu m}$ \\
    Particle density, $\rho_p$ & $5400,22500$ & $\mathrm{kg/m^3}$ \\
    Particle charge, $q$ & $\{0,0.5,1\} \times 10^{-14}$ & $\mathrm{C}$ \\
    Particle number, $N_p$ & $5 \times 10^4$ & $-$\\
    Domain-averaged particle volume fraction, $\overline{\alpha}$ & $1.33 \times 10^{-6}$ & $-$ \\

  \end{tabular}
  \caption{Simulation parameters.}
  \label{tab:SimulationParameters}
  \end{center}
\end{table}

The total number of particles in the domain is $N_p=5 \times 10^4$, and the particles are assumed to be heavy and small. The particle diameter is fixed at $d_p=20 \ \mathrm{\mu m}$ ($d_p^+=0.36$), so the domain-averaged particle volume fraction is a constant ($\overline{\alpha}=1.33 \times 10^{-6}$) and falls within the dilute regime. The particle Stokes number is controlled by adjusting the particle density. 

\subsection{Fluid phase}
\label{sec:Fluid phase}
In this study, the volume-filtered Eulerian-Lagrangian framework is employed to simulate particle-laden turbulent channel flow. The incompressible fluid motion is solved using the open-source solver NGA2 \citep{DesjardinsJCP2008, CapecelatroJCP2013}. A brief derivation of the volume-filtered governing equation for the fluid phase, starting from the standard point-wise equations, is provided in Appendix \ref{sec:Appendix_VFEL}.1. The associated model closure problem is further discussed in Appendix \ref{sec:Appendix_VFEL}.2. Finally, the volume-filtered governing equations of the fluid phase used in this study are given by

\begin{subequations}
\label{eq:NS equation}
\begin{equation}
    \frac{\partial{\alpha}}{\partial t} + \nabla \cdot (\alpha \mathbf{u}_f) = 0, 
\end{equation}

\begin{equation}
    \frac{\partial (\alpha \mathbf{u}_f)}{\partial t} + \nabla \cdot (\alpha \mathbf{u}_f \otimes \mathbf{u}_f) = \nabla \cdot \mathbf{\tau} + \mathbf{f}_{F} + \mathbf{f}_{P}.
\end{equation}

\noindent
Here $\alpha$ and $\mathbf{u}_f$ are the fluid volume fraction and the flow velocity. The fluid stress is $\mathrm{\tau}=-p/\rho_f \mathbf{I} + \nu_f [(\nabla \mathrm{u}_f+{\nabla \mathrm{u}_f}^T) - 2(\nabla \cdot \mathbf{u}_f)\mathbf{I}/3]$ with $p$, $\rho_f$, $\nu_f$ being the pressure, density, and kinematic viscosity of the fluid phase, respectively. $\mathbf{I}$ is the identity tensor. $\mathbf{f}_{F}$ is the streamwise forcing term that maintains a constant mass flow rate. $\mathbf{f}_{P}$ is the momentum exchange term due to inter-phase coupling. 

The volume-filtered Navier-Stokes equations are solved on a staggered grid with second-order spatial accuracy for both the convective and the viscous term, and are advanced using the second-order semi-implicit Crank-Nicolson scheme \citep{PierceBook2001}. The pressure Poisson equation is solved by a multigrid solver using the preconditioned conjugate gradient method \citep{FalgoutICCS2002}.

\end{subequations}

\subsection{Particle phase}
The suspended particles are treated as spheres and their movements are simulated using the Lagrangian approach. Both particle translation and rotation are updated considering the exerted forces/torques as

\begin{subequations}
\label{eq:DEM}
    \begin{equation}
        m_i \frac{\mathrm{d}\mathbf{v}_i}{\mathrm{d}t} = \mathbf{F}_i^{F} + \mathbf{F}_i^{C} + \mathbf{F}_i^{E},
    \end{equation}

    \begin{equation}
        I_i \frac{\mathrm{d}\mathbf{\Omega}_i}{\mathrm{d}t} = \mathbf{T}_i^{F} + \mathbf{T}_i^{C}.
    \end{equation}
\end{subequations}
 
\noindent
Here $m_i= \pi \rho_p d_{\mathrm{p},i}^3/6$ and $I=m_i d_{\mathrm{p},i}^2/10$ are the mass and the momentum of inertia of particle $i$. $\mathbf{v}_i$ is the particle velocity, $\mathbf{\Omega}_i$ is the rotation rate, and $\mathbf{F}_i$ and $\mathbf{T}_i$ denote the acted force and torque. The superscripts $F$, $C$ and $E$ refer to fluid force/torque, collision force/torque and electrostatic force, respectively. 

In this study, gravity is intentionally neglected. The presence of wall-normal gravity would introduce an additional vertical migration velocity, increasing particle flux towards the bottom wall and decreasing it towards the top wall \citep{MarchioliIJMF2007, BerkJFM2020}. In contrast, as will be shown below, both the turbophoresis effect and the electrostatic force tend to enhance particle deposition towards both walls. Consequently, incorporating gravity could break the symmetry of the system, with the steady-state statistics being governed by a complex interplay between gravity, electrostatics, and particle-turbulence interactions. This added complexity could make it more challenging to isolate and clarify the specific role of electrostatic forces. For this reason, we have intentionally neglected gravity, ensuring that any changes in particle dynamics between neutral and charged cases can be solely attributed to the influence of electrostatic forces.

\subsubsection{Particle-fluid interaction}
\label{sec:FluidForce}
The particles considered in this study are significantly heavier than the fluid ($\rho_p/\rho_f \sim O(10^3)$), and their size is small compared to the viscous length ($d_p/\delta_\nu$=0.36). Given that the length scales of near-wall turbulent structures are at least tens of $\delta_\nu$, the solid particles can be treated as point particles. 

For an individual particle $i$, the full fluid force can be obtained by integrating the fluid stress over the particle surface. As the volume-filtered framework is used, the fluid force can be decomposed into the contribution from the resolved and unsolved stress as

\begin{equation*}
    \mathbf{F}_i^F=\int_{\mathcal{S}_i}\mathbf{\tau}\cdot \mathbf{n} \mathrm{d}\mathbf{y}=\int_{\mathcal{S}_i} (\overline{\mathbf{\tau}} + \mathbf{\tau}^\prime) \cdot \mathbf{n} \mathrm{d}\mathbf{y} = \int_{\mathcal{V}_i} \nabla \cdot \overline{\mathbf{\tau}} \mathrm{d}\mathbf{y} + \int_{\mathcal{S}_i} \mathbf{\tau}^\prime \cdot \mathbf{n} \mathrm{d}\mathbf{y}.
\end{equation*}

If the particle size is much smaller than the filter size, as in this study, $\nabla \cdot \overline{\mathbf{\tau}}$ varies little at the particle scale and can be taken out of the integral. The fluid force then becomes

\begin{equation*}
    \mathbf{F}_i^F=\int_{\mathcal{V}_i} \nabla \cdot \overline{\mathbf{\tau}} \mathrm{d}\mathbf{y} + \int_{\mathcal{S}_i} \mathbf{\tau}^\prime \cdot \mathbf{n} \mathrm{d}\mathbf{y} \approx \nabla \cdot \overline{\mathbf{\tau}} \mathcal{V}_i +\int_{\mathcal{S}_i} \mathbf{\tau}^\prime \cdot \mathbf{n} \mathrm{d}\mathbf{y}.
\end{equation*}

\noindent
Here $\mathcal{V}_i$ is the volume of particle $i$. The fluid force due to the residual stress, $\int_{\mathcal{S}_i} \mathbf{\tau}^\prime \cdot \mathbf{n} \mathrm{d}\mathbf{y}$, needs to be modeled. As discussed in Appendix \ref{sec:Appendix_VFEL}, the eddy viscosity at the unresolved scale, $\nu_t$, is much smaller than the fluid molecular viscosity, $\nu_f$, indicating that the unresolved flow around the particle is essentially laminar. Based on these considerations, the fluid force is modeled using the Maxey-Riley equation \citep{MaxeyPOF1983}. Since the fluid drag is the dominant fluid force, other fluid forces are neglected. A detailed comparison of the fluid drag with other forces, such as lift force and short-range lubrication force, is provided in Appendix \ref{sec:Appd_FluidForce}. The resolved fluid force, $\nabla \cdot \overline{\mathbf{\tau}} \mathcal{V}_i$, is also negligible compared to fluid drag for two reasons. First, the filter size $\delta_F$ is much larger than the particle size $d_p$, resulting in a small divergence of the filtered stress. Second, the particle size $d_p$ is small, leading to an even smaller volume $\mathcal{V}$. Preliminary tests show that the ratio of the resolved fluid force to the drag force, $|\nabla \cdot \overline{\mathbf{\tau}} \mathcal{V}|/F_{d}$, is only 0.036 for $St^+=32$ and 0.001 for $St^+=133$. Consequently, we only consider fluid drag in this study, and the fluid force and torque are given as

\begin{subequations}
    \begin{equation}
        \mathbf{F}_i^F = -3 \pi \mu_f d_{p,i} \left(\mathbf{v}_i-\mathbf{u}_f(\mathbf{x}_i)\right) f_I,
    \end{equation}

    \begin{equation}
        \mathbf{T}_i^F = -\pi \mu_f d_{p,i}^3 \left(\mathbf{\Omega}_i-\frac{1}{2} \mathbf{\omega}(\mathbf{x}_i)\right).
    \end{equation}
\end{subequations}

\noindent
Here $\mu_f$ is the fluid dynamic viscosity, $\mathbf{u}_f(\mathbf{x}_i)$ and $\mathbf{\omega}(\mathbf{x}_i)$ are the fluid velocity and vorticity interpolated at the particle location using trilinear interpolation. The influence of the order of the interpolation scheme has been discussed in Appendix \ref{sec:Appd_Interpolation}. In two-way coupled simulations, the accurate calculation of fluid drag requires the undisturbed fluid velocity $\Tilde{\mathbf{u}}_f(\mathbf{x}_p)$ at the particle location, because the feedback force from the target particle itself perturbs surrounding fluid flow. As a result, the local fluid velocity, $\mathbf{u}_f(\mathbf{x}_p) (\neq \Tilde{\mathbf{u}}_f(\mathbf{x}_p))$, is effectively disturbed (or `contaminated'), leading to an underestimated slip velocity and, consequently, a reduced drag force. To address this issue, various correction schemes have been proposed for both point-particle \citep{GualtieriJFM2015, HorwitzPRF2020} and finite-size particle simulations \citep{BalachandarIJMF2023} to recover the undisturbed fluid velocity $\Tilde{u}_f(x_p)$ and ensure physically accurate results. In this work, however, because of the large size ratio between the Gaussian filter length and the particle size $\delta_F/d_p=8$, the error in drag force caused by self-induced disturbance is less significant, so the correction scheme is not applied. Detailed discussions on the correction scheme of the undisturbed fluid velocity and its influences are given in Appendix \ref{sec:Appd_Undisturbed}. To account for the effect of fluid inertia, the drag force is corrected using the Schiller–Naumann correction factor, $f_I$, which writes

\begin{equation}
    f_I=1+0.15 Re_p^{0.687}.
\end{equation}

\noindent
Here the particle Reynolds number is defined as $Re_p=|\mathbf{v}_i-\mathbf{u}_f(\mathbf{x}_i)| d_{p,i}/\nu_f$.

To consider the flow modulation caused by the particle phase, both the fluid volume fraction $\alpha$ and the momentum transfer term $\mathbf{f}_P$ in (\ref{eq:NS equation}) are computed as follows

\begin{subequations}
    \begin{equation}
    \label{eq:GaussFilter1}
        \alpha(\mathbf{X}_i) = 1- \frac{1}{V_{cell,i}} \sum_{j=1}^{N_p} G_F(|\mathbf{X}_i-\mathbf{x}_j|) V_{p,j},
    \end{equation}

    \begin{equation}
    \label{eq:GaussFilter2}
        \mathbf{f}_P (\mathbf{X}_i) = -\frac{1}{\rho_f V_{cell,i}} \sum_{j=1}^{N_p} G_F(|\mathbf{X}_i-\mathbf{x}_j|) \mathbf{F}^F_{j}.
    \end{equation}
\end{subequations}

\noindent
Here $\mathbf{X}_i$ is the location of the $i$th grid cell, $V_{p,j}=\pi d_{p,j}^3 /6$ is the volume of the $j$th particle. $G_F$ is the fluid Gaussian filter that distributes the Lagrangian quantities (i.e., $V_{p,j}$ and $\mathbf{F}^F_{j}$) to the Cartesian mesh. The characteristic fluid filtering length $\delta_F$, defined as the full width of the fluid Gaussian filter $G_F$ at the half height, is chosen as $\delta_F=8d_p=2.88 \delta_\nu$ so that the turbulent structures are sufficiently resolved \citep{CapecelatroJFM2014}.

\subsubsection{Particle-particle collision}
\label{sec:CollisionForce}
If the center-to-center distance between a pair of particles $i$ and $j$ is smaller than the sum of their radii ($|\mathbf{x}_i-\mathbf{x}_j|<(d_{p,i}+d_{p,j})/2$), these particles are in contact, and the collision forces and torque are considered. The contact force from particle $j$ to $i$ is given by

\begin{equation}
    \mathbf{F}^C_{i \leftarrow j} = F_n \mathbf{n} + F_t \mathbf{t},
\end{equation}

\noindent
where $\mathbf{n}=(\mathbf{x}_j-\mathbf{x}_i)/|\mathbf{x}_j-\mathbf{x}_i|$ is the unit vector pointing from the centroid of particle $i$ to that of particle $j$, and the tangent unit vector $\mathbf{t}=\mathbf{v}_{rel,t}/|\mathbf{v}_{rel,t}|$ follows the tangential relative velocity $\mathbf{v}_{rel,t}$ at the contact point. The contact force components are given by

\begin{subequations}
    \begin{equation}
        F_n = -k_n \delta_n,
    \end{equation}

    \begin{equation}
        F_t = -\mu_t |\mathbf{F}_n|.
    \end{equation}
\end{subequations}

\noindent
The normal force $F_n$ follows the Hertzian contact theory and accounts for the elastic repulsion between contact particles. The normal overlap is $\delta_n = (d_{p,i}+d_{p,j})/2-|\mathbf{x}_i-\mathbf{x}_j|$, and the normal elastic stiffness can be expressed as $k_n = 4E \sqrt{R \delta_n}/3$. Here $R=(1/r_i+1/r_j)^{-1}$ is the effective radius, and $E=((1-\nu_{p,i}^2/E_i) + (1-\nu_{p,j}^2/E_j))^{-1}$ is the effective elastic modulus. $r_i$, $E_i$ and $\nu_{p,i}$ are the radius, Young's modulus and the Poisson ratio of particle $i$, respectively. The tangent force $F_t$ is determined from the static friction model with the friction coefficient $\mu_t =0.3$ chosen based on experimental measurements \citep{ThorntonPT1991}. The associated torque is then determined as 

\begin{equation}
    \mathbf{T}_{i \leftarrow j}^C = \mathbf{r}_{C,ij} \times (F_t \mathbf{t}).
\end{equation}

\noindent
Here $\mathbf{r}_{C,ij}$ points from the center of particle $i$ to the contact point between $i$ and $j$. Once the collision force and torque from each contact neighbor $j$ is computed, the total collision force and torque in (\ref{eq:DEM}) can be obtained as $\mathbf{F}_i^C = \sum_j \mathbf{F}^C_{i \leftarrow j}$ and $\mathbf{T}_i^C = \sum_j \mathbf{T}^C_{i \leftarrow j}$. Note that the collision interactions between a particle and a wall can be computed similarly by treating the wall as a particle at rest with infinite radius and mass.

\subsection{Validation of neutral particle-laden simulations}
Several cases presented in figure 14 of \citet{JohnsonJFM2020} are selected as benchmark results to validate our solvers for the particle-laden turbulent flows. The key parameters for these cases are summarized in table \ref{tab:ValidationParameters}.

In the reference, a standard two-way coupled Eulerian-Lagrangian framework is employed to simulate the turbophoresis of small inertial particles in a turbulent channel flow. The channel flow is resolved using a grid number of $172 \times 86 \times 128$, and the friction Reynolds number is $\mathrm{Re}_{\tau}=150$. Neutral solid particles are subject to fluid drag force, while interparticle collisions are modeled using a hard-sphere model. A restitution coefficient of $e=1.0$ is used, indicating that collisions are purely elastic. The effects of two-way coupling are also accounted for. 

\begin{table}
  \begin{center}
\def~{\hphantom{0}}
  \begin{tabular}{lll}
    Dataset & figure 14 \citep{JohnsonJFM2020} & Our simulation \\
    \\
    \textit{Fluid phase} \\
    Domain size, $L_x \times L_y \times L_z$ & $4 \pi \delta \times 2 \delta \times 2 \pi \delta $ & $4 \pi \delta \times 2 \delta \times 2 \pi \delta $\\
    Grid number, $N_x \times N_y \times N_z $ & $172 \times 86 \times 128$ & $128^3$ \\
    Friction Reynolds number, $\mathrm{Re}_{\tau}$ & $150$ & $150$ \\
    \\
    \textit{Particle phase} \\
    Particle diameter, $d_p^+$ & $0.5$ & $0.5$ \\
    Particle Stokes number, $St^+$ & $32,128$ & $32,128$ \\
    Particle volume fraction, $\overline{\alpha}$ & $3 \times 10^{-6} - 1 \times 10^{-4}$ & $3 \times 10^{-6} - 1 \times 10^{-4}$ \\
    Fluid force & Drag & Drag\\
    Collision model & Hard sphere & Soft sphere\\
    Restitution coefficient, $e$ & $1.0$ & $1.0$\\
    Interphase coupling & Two-way coupled & Two-way coupled\\
  \end{tabular}
  \caption{Parameters in validation cases.}
  \label{tab:ValidationParameters}
  \end{center}
\end{table}

In our simulation, both the domain size and the Reynolds number are chosen to match those in the reference, while the grid resolution of $128^3$ is consistent with that introduced in section \ref{sec:system}. In the reference, simulations were conducted for four different Stokes numbers ($St^+=1, 32, 128, 512$). However, we validate the results only for $St^+=32$ and $128$, as these values are more relevant to the particle inertia discussed in this study. The particle diameter is fixed at $d_p=0.5 \delta_\nu$, and the particle densities are set to $\rho_p = 2765 \ \mathrm{kg/m^3}$ ($St^+=32$) and $ 11059 \ \mathrm{kg/m^3}$ ($St^+=128$) to achieve the desired Stokes number. The numbers of particles in the simulations vary according to different particle volume fractions: $N_p=24429$ ($\overline{\alpha}_p=3 \times 10^{-6}$), $N_p=81430$ ($\overline{\alpha}_p=1 \times 10^{-5}$), $N_p=244290$ ($\overline{\alpha}_p=3 \times 10^{-5}$), $N_p=814300$ ($\overline{\alpha}_p=1 \times 10^{-4}$). The drag force is computed as described in section \ref{sec:FluidForce}, while the normal collision force is resolved using the soft-sphere Hertzian contact model (section \ref{sec:CollisionForce}), assuming that collisions are elastic. Finally, interphase coupling is incorporated following the approach mentioned in section \ref{sec:FluidForce}.

\begin{figure}
    \centering
    \includegraphics[width=13.5cm]{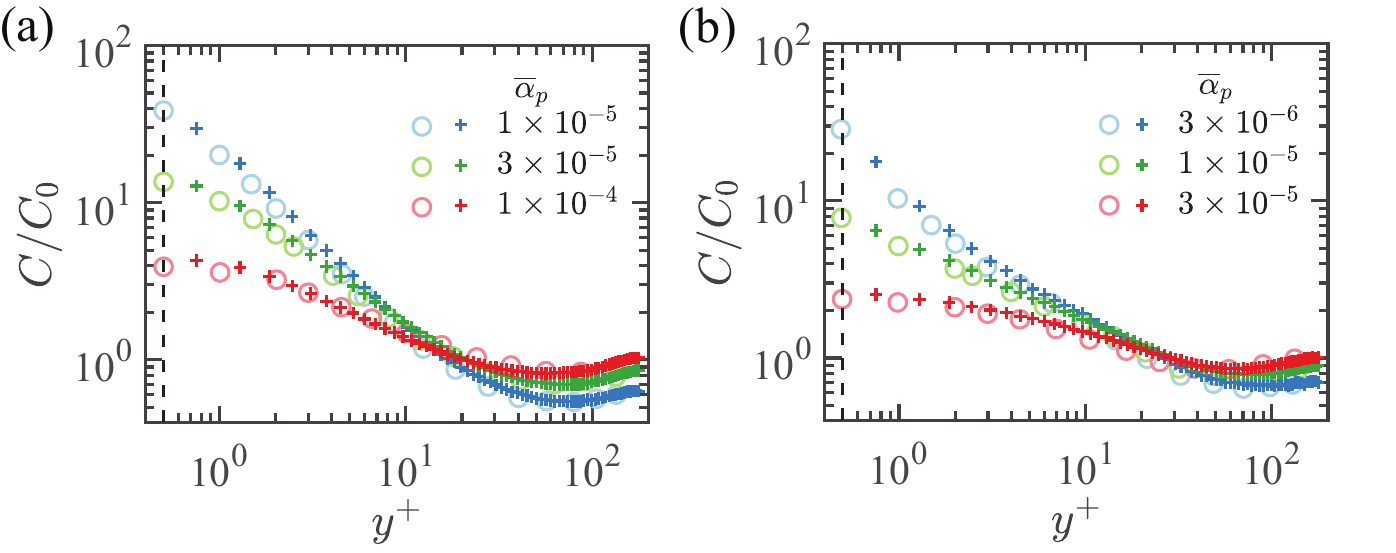}
    \caption{Steady wall-normal particle concentration $C/C_0$ for particles with (a) $St^+=32$ and (b) $St^+=128$. Circles ($\circ$) denote profiles obtained from \citet{JohnsonJFM2020}, while plus signs ($+$) represent simulation results using the methods introduced in this study.}
    \label{fig:Validation_Concentration}
\end{figure}

Figure \ref{fig:Validation_Concentration} compares the wall-normal particle concentration profiles in the steady state. The vertical dashed line ($y^+=0.5$) marks the location where particles collide with the wall. The profiles corresponding to different $St^+$ and $\overline{\alpha}_p$ show reasonable agreement, demonstrating the reliability of both the fluid and particle solvers.

\subsection{Electrostatic interaction}
\subsubsection{Particle-particle-particle-mesh method}
\label{sec:P3M}
The particle-particle-particle-mesh method (P$^3$M) is employed to calculate the Eulerian electric field and to resolve the electrostatic interaction acting on charged particles \citep{HockneyBook2021, YaoPRF2018}. The particle charges are assumed as point charges located at particle centers, and the electrostatic force acting on particle $i$ is

\begin{equation}
\mathbf{F}_{i}^E = q_i \mathbf{E}(\mathbf{x}_i),
\end{equation}

\noindent
where $q_i$ is the particle charge and $\mathbf{E}(\mathbf{x}_i)$ is the electric field at the particle location $\mathbf{x}_i$. The idea of P$^3$M is to split the electrostatic field into two parts: 

\begin{equation}
\mathbf{E}(\mathbf{x}_i)=\mathbf{E}_M(\mathbf{x}_i)+\mathbf{E}_C (\mathbf{x}_i).
\end{equation}

\noindent
Here $\mathbf{E}_M (\mathbf{x}_i)$ is the long-range contribution that can be efficiently obtained from the Eulerian electric field, while $\mathbf{E}_C (\mathbf{x}_i)$ is the short-range correction that only needs to be included when other particles are within a critical distance $r_{cut}$ to the target particle.

To find the long-range contribution $\mathbf{E}_M (\mathbf{x}_i)$, the point charges $q_j$ carried by discrete particles located at $\mathbf{x}_j$ are first filtered and sent to the Cartesian mesh. The resulting volumetric charge density $\rho_M$ on the mesh is

 \begin{equation}
 \label{eq:ChargeFilter}
     \rho_M (\mathbf{X}_i) = \frac{1}{V_{cell,i}}\sum_{j} q_j G_E(|\mathbf{X}_i-\mathbf{x}_j|),
 \end{equation}

\noindent
where the electric Gaussian filter is

\begin{equation}
    G_E(\mathbf{r})=\frac{\beta^3}{\pi^{3/2}} e^{-\beta^2 |\mathbf{r}|^2}.
\end{equation}

\noindent
The width of the Gaussian filter at the half height is related to $\beta$ by $\delta_E=2\sqrt{2\ln{2}}/ \beta$. The electric Poisson equation (\ref{eq:ES_Poisson_Eq}) is discretized to the second-order spatial accuracy, and is solved for the electric potential $\phi_M$ using the same method as that for the pressure Poisson equation in Sec. \ref{sec:Fluid phase}. The electric field ($\mathbf{E}_M$) is then determined by (\ref{eq:E_Field_Eq}) with the fourth-order central differencing scheme. Finally, the electric field at the particle locations ($\mathbf{E}_M (\mathbf{x}_i)$) is further interpolated using the fourth-order Lagrange interpolation.

\begin{subequations}
    \begin{equation}
    \label{eq:ES_Poisson_Eq}
    \nabla^2 \phi_{\mathrm{M}} = -\frac{\rho_M}{\epsilon_0},
    \end{equation}

    \begin{equation}
    \label{eq:E_Field_Eq}
    \mathbf{E}_M=-\nabla \phi_M.
    \end{equation}
\end{subequations}

For a particle $j$ at $\mathbf{x}_j$ that is close to the target particle $i$ at $\mathbf{x}_i$, the filtered field contribution using (\ref{eq:ChargeFilter}) to (\ref{eq:E_Field_Eq}) is

\begin{equation}
\label{eq:Filtered_E_Field}
    \mathbf{E}_{M,ij} = \frac{q_j \mathbf{r}_{ij}}{4 \pi \varepsilon_0 |\mathbf{r}_{ij}|^3} \left[ 1- \mathrm{erfc} \left(\beta |\mathbf{r}_{ij}|\right) - \frac{2 \beta |\mathbf{r}_{ij}|}{\sqrt{\pi}}\mathrm{exp} \left(-\beta^2 |\mathbf{r}_{ij}|^2 \right) \right],
\end{equation}

\noindent
where $\mathbf{r}_{ij} = \mathbf{x}_i-\mathbf{x}_j$ is the vector pointing from $\mathbf{x}_j$ to $\mathbf{x}_i$, and erfc is the complimentary error function. Meanwhile, the exact contribution should be

\begin{equation}
\label{eq:Exact_E_Field}
    \mathbf{E}_{exact,ij} = \frac{q_j \mathbf{r}_{ij}}{4 \pi \varepsilon_0 |\mathbf{r}_{ij}|^3}.
\end{equation}

\noindent
To eliminate the error due to filtering, the short-range correction is added if the interparticle distance is within the cut-off distance $r_{cut}$ as

\begin{equation}
\label{eq:ES_ShortCorrect}
\begin{split}
    \mathbf{E}_C (\mathbf{x}_i) & = \sum_{ \substack{j \neq i \\ |\mathbf{r}_{ij}|<r_{cut}} } \left( \mathbf{E}_{exact,ij} - \mathbf{E}_{M,ij} \right) \\
    & = \sum_{\substack{j \neq i \\ |\mathbf{r}_{ij}|<r_{cut}}} \frac{q_j \mathbf{r}_{ij}}{4 \pi \varepsilon_0 |\mathbf{r}_{ij}|^3} \left[ \mathrm{erfc} \left(\beta |\mathbf{r}_{ij}|\right) + \frac{2 \beta |\mathbf{r}_{ij}|}{\sqrt{\pi}} \mathrm{exp}\left(-\beta^2 |\mathbf{r}_{ij}|^2 \right) \right].
\end{split}
\end{equation}

\begin{figure}
    \centering
    \includegraphics[width=13.5cm]{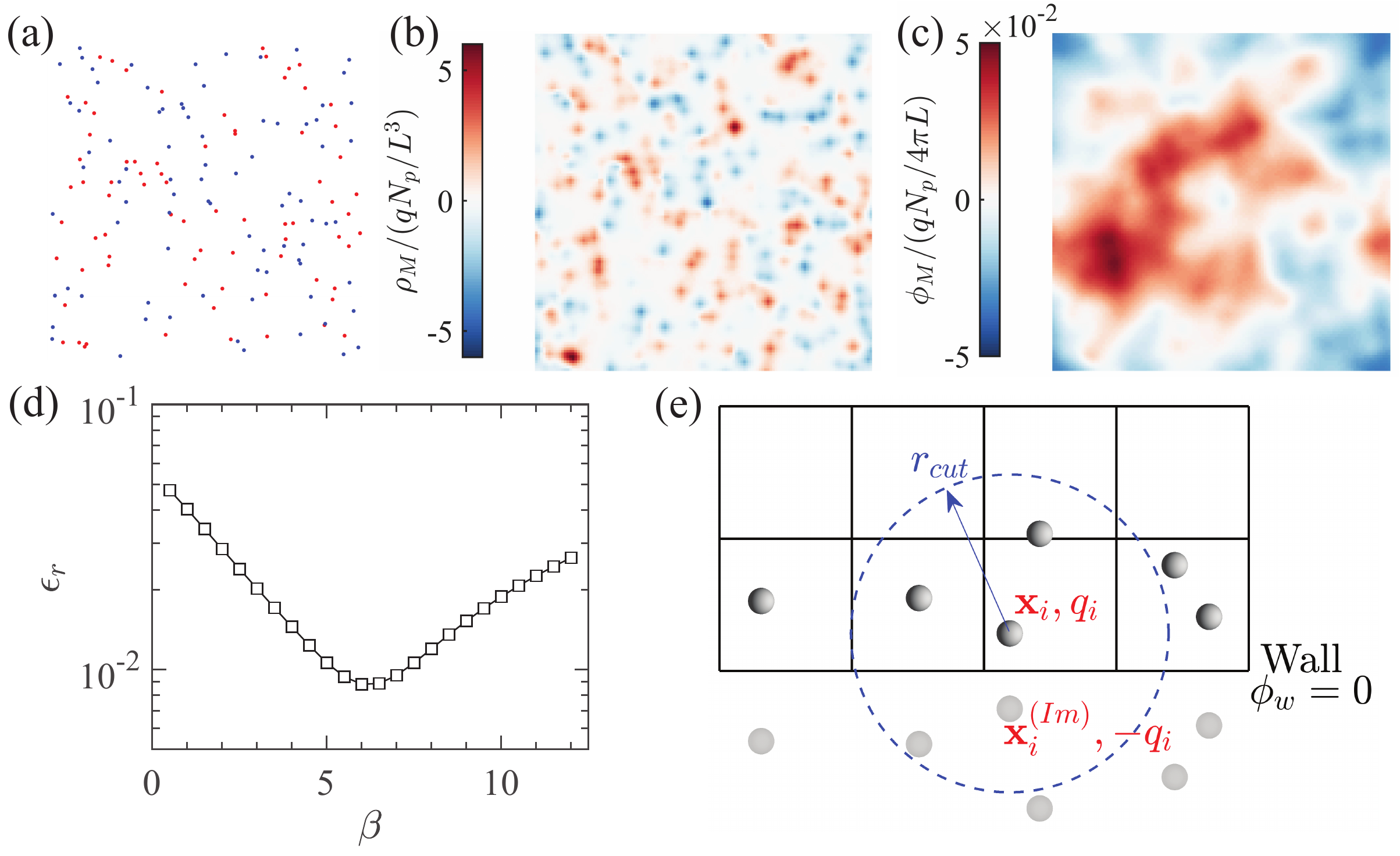}
    \caption{Schematic of the P$^3$M validation: (a) positive/negative (red/blue) point charges carried by particles, (b) the normalized charging density $\rho_M / (q N_p/L^3)$, and (c) the normalized electric potential $\phi_M / (q N_p/4 \pi L)$ in a thin slice. (d) Dependence of the relative error $\epsilon_r$ (\ref{eq:ElectricError}) of P$^3$M method on the parameter $\beta$. (e) Dirichlet boundary conditions at the wall ($\phi_w=0$) and the added image particles.}
    \label{fig:EsMethod}
\end{figure}

To validate the accuracy of the P$^3$M method, the electrostatic forces calculated from both the P$^3$M method and the standard Ewald summation \citep{DesernoJChemPhys1998} are compared. Details about the Ewald summation are introduced in Appendix \ref{sec:Ewald}. In the test case, $N_p=5000$ particles are randomly placed in a triply periodic domain with the side length $L=2\pi$. Half of the particles carry a nominal positive charge $q=1$ while the others carry a nominal negative charge $q=-1$ (Fig. \ref{fig:EsMethod}(a)-(c)). When implementing P$^3$M, the Cartesian grid number is set to $128^3$. The cut-off distance is fixed at $r_{cut}=0.2$ for the following reasons. First, $r_{cut}$ needs to be sufficiently large to ensure the convergence of short-range corrections for all particles. At the same time, $r_{cut}$ cannot be too large, as this would significantly increase computational cost. In the test case, for a fixed $\beta$, $r_{cut}$ is gradually increased, and the normal of the residual electrostatic force, $|\mathbf{F}^E-\mathbf{F}^{E,Ewald}|$ (the numerator in (\ref{eq:ElectricError})), is calculated. As $r_{cut}$ increases, the residual force continues to decrease and approaches a minimum at around $r_{cut}=0.2$. Based on this result, $r_{cut}=0.2$ is selected for the test, which ensures both force convergence and computational efficiency. The value of $\beta$ is then swept to change the electric filter length $\delta_E$. The P$^3$M results are denoted by $\mathbf{F}_i^E$, and the relative error $\epsilon_r$ is calculated by

\begin{equation}
\label{eq:ElectricError}
    \epsilon_r = \frac{|\boldsymbol{F}^E-\boldsymbol{F}^{E,Ewald}|}{|\boldsymbol{F}^{E,Ewald}|} =\frac{[ \sum_{i=1}^{N_\mathrm{p}} (\boldsymbol{F}_i^E-\boldsymbol{F}_i^{E,Ewald})^2 /N_{\mathrm{p}}]^{1/2}}{[ \sum_{i=1}^{N_\mathrm{p}} (\boldsymbol{F}_i^{E,Ewald})^2 /N_{\mathrm{p}}]^{1/2}},
\end{equation}

\noindent
The dependence of $\epsilon_r$ on $\beta$ is shown in Fig.\ref{fig:EsMethod}(d). The relative error reaches the minimum $\epsilon_r=0.88\%$ at $\beta=6.0$, thus verifying the reliability of the P$^3$M method.

\subsubsection{Electrical boundary conditions}
In the channel, both the top and bottom boundaries are assumed to be grounded conductive walls. When solving the electric Poisson equation (\ref{eq:ES_Poisson_Eq}), periodic boundary conditions are applied in the streamwise ($x$) and the spanwise ($z$) directions, and zero-Dirichlet boundary conditions are added at both walls ($y=\pm \delta$):

\begin{equation}
\label{eq:ES_Dirichlet}
    \phi_w=0.
\end{equation}

\noindent
Note that (\ref{eq:ES_Dirichlet}) only ensures an appropriate electrical boundary condition on the mesh. When charged particles are close to the wall, the length scale of the local electric field is usually much smaller than the cell size and cannot be fully resolved. Therefore, image particles are added to consider such near-wall effects \citep{LiuIJNME2010, YaoPT2021}. If the distance between a particle $i$ and the wall is smaller than $r_{cut}$, its image is added at the symmetric location $\mathbf{x}_i^{(Im)}$ about the wall with opposite polarity $q_i^{(Im)}=-q_i$. When summing the short-range correction force in (\ref{eq:ES_ShortCorrect}), the contribution of all the image particles within $r_{cut}$ is also added (Fig.\ref{fig:EsMethod}(e)):

\begin{equation}
\label{eq:ES_Image}
    \mathbf{E}_C^{(Im)}(\mathbf{x}_i) = \sum_{|\mathbf{r}_{ij}^{(Im)}|<r_{cut}} \frac{q_j^{(Im)} \mathbf{r}_{ij}^{(Im)}}{4 \pi \varepsilon_0 |\mathbf{r}_{ij}^{(Im)}|^3} \left[ \mathrm{erfc} \left(\beta |\mathbf{r}_{ij}^{(Im)}|\right) + \frac{2 \beta |\mathbf{r}_{ij}^{(Im)}|}{\sqrt{\pi}} \mathrm{exp}\left(-\beta^2 |\mathbf{r}_{ij}^{(Im)}|^2 \right) \right].
\end{equation}

\noindent
Here $\mathbf{r}_{ij}^{(Im)}$ points from the image of particle $j$ to the target particle $i$. Therefore, the near-wall correction can be taken as a special case of the short-range correction (\ref{eq:ES_ShortCorrect}) due to all the images. 

Furthermore, to avoid over-filtering the electric field, the electric filter length is chosen to be $\delta_E=5 \delta_\nu$ in the simulations. The cut-off distance $r_{cut}=36 \delta_\nu$ is set larger than $\delta_E$ so that the short-range correction is converged.

We now note that, using P$^3$M, the accuracy of the particle-wall (PW) electrostatic force is inherently equivalent to that of the PP electrostatic force. When evaluating the electric field $\mathbf{E}(\mathbf{x}_i)$ at the particle locations in a wall-bounded domain, the conducting wall can influence both the long-range contribution, $\mathbf{E}_M(\mathbf{x}_i)$, and the short-range correction, $\mathbf{E}_C(\mathbf{x}_i)$. First, the electric Poisson equation is solved using periodic boundary conditions in the $x$ and $z$ directions, and a zero-Dirichlet boundary condition ($\phi_w=0$) at the walls. Since the same Poisson solver is employed with an identical tolerance of $\epsilon_{tol}=10^{-6}$, the accuracy of the electric field on the mesh, $\mathbf{E}_M$, in the wall-bounded case is comparable to that in the triply periodic case. The electric field at particle locations is then interpolated using the same fourth-order Lagrangian interpolation, ensuring that the long-range contribution, $\mathbf{E}_M(\mathbf{x}_i)$, remains equally accurate in the channel. For the short-range contribution from image particles, the short-range correction for image particles (\ref{eq:ES_Image}) has the same functional form as that for real particles (\ref{eq:ES_ShortCorrect}). The only difference lies in the positions and charges of the image particles. Therefore, when summing the short-range corrections within the same cut-off distance, $r_{cut}$, contributions from both real and image particles are calculated together. This approach guarantees that the accuracy of $\mathbf{E}_C(\mathbf{x}_i)$ is preserved. Consequently, the accuracy of P$^3$M in a wall-bounded domain is of the same order as in a triply periodic domain.

\section{Results and discussions}

\subsection{Wall-normal transport and deposition velocity of charged particles}
\label{sec:Deposition}

In each case, particles are released into a fully developed turbulent flow with random initial positions and zero velocity. The particle spatial distribution then starts to evolve from the initially random state towards a steady state that is characterized by a high concentration near the wall. To quantify the temporal evolution of the particle phase, the Shannon entropy $\mathcal{S}$ is used to describe the non-uniformity of the wall-normal particle distribution \citep{PicanoPOF2009, SardinaJFM2012}. It takes approximately $(1-2) \times 10^4 \tau_{\nu}$ for the particle distribution to transition from the initial random distribution to a steady state, where $\mathcal{S}$ is independent of time (not shown). Statistics are then taken over another $5 \times 10^3 \tau_{\nu}$ and presented below. However, for the case with moderate inertia ($St^+=32$) and the highest charge ($q=1 \times 10^{-14} \ \mathrm{C}$), a steady state was not reached after a simulation period exceeding $2 \times 10^4 \tau_{\nu}$. This case is thus excluded from the current discussion of steady-state statistics.

\begin{figure}
    \centering
    \includegraphics[width=13.5cm]{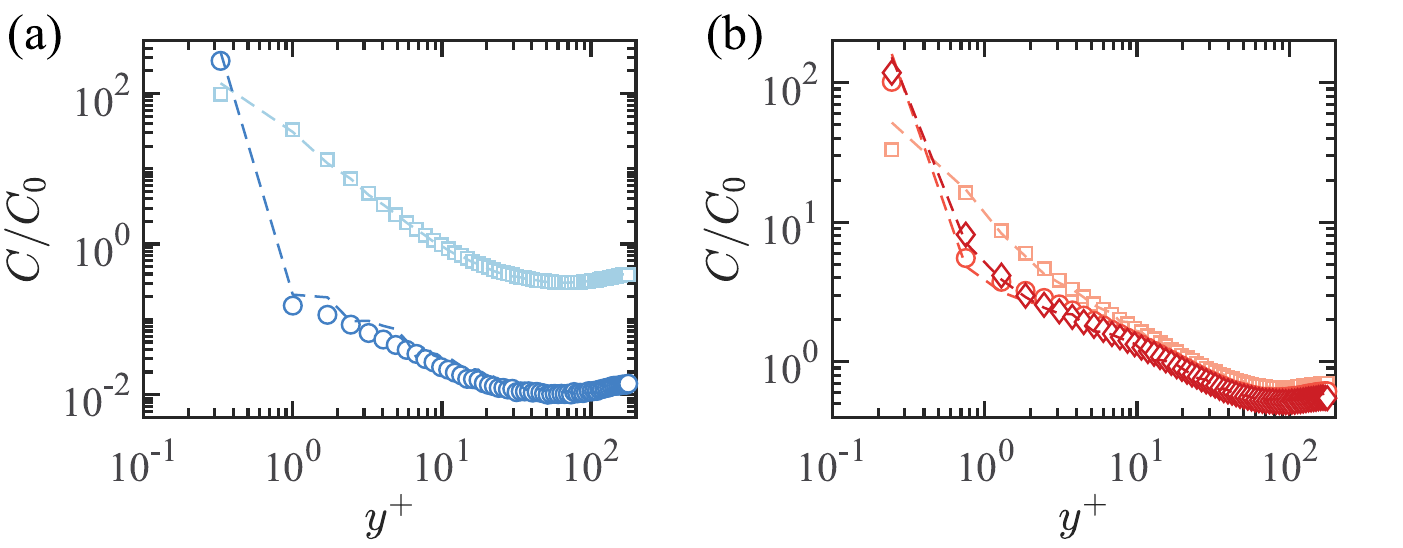}
    \caption{Normalized wall-normal particle concentration $C/C_0$ for both neutral and charged particles with (a) $St^+=32$ and (b) $St^+=133$. Scatters are simulation results and dashed lines are predictions using (\ref{eq:Concentration}). Colors from light to dark represent results for $q=0 \ \mathrm{C}$, $5 \times 10^{-15} \ \mathrm{C}$, and $1 \times 10^{-14} \ \mathrm{C}$.}
    \label{fig:Concentration}
\end{figure}

We start with the distribution of charged particles in the wall-normal direction. Figure \ref{fig:Concentration} compares the normalized wall-normal particle concentration $C/C_0$ between neutral and charged particles. The local particle concentration $C(y)$ is equal to the number of particles in each wall-normal bin divided by the bin volume, and the average concentration is $C_0=N_p/(L_x L_y L_z)$. In figure \ref{fig:Concentration}, the simulation results are represented by scatters, while the dashed lines are model predictions based on (\ref{eq:Concentration}) that will be further detailed in section \ref{sec:Model}. In the neutral cases, particles driven by turbophoresis migrate from the outer layer towards the wall, leading to the increase of $C(y)$ as $y^+$ decreases. Compared with the more inertial particles with $St^+=133$, particles with $St^+=32$ show more pronounced accumulation ($C(y^+=1)/C_0 \sim 10^2$) in the neutral case as these particles are more responsive to near-wall coherent structures. 

Once particles are charged, the image force further attracts particles towards the walls leading to more significant accumulation. As shown in figure \ref{fig:Concentration}(a), most particles with $St^+=32$ and $q=5 \times 10^{-15} \ \mathrm{C}$ remain concentrated in the innermost bin within the viscous layer, while their concentration in both the buffer layer and the outer layer is drastically reduced. A similar trend is observed for particles with $St^+=133$, though to a lesser extent due to their larger inertia. For $St^+=133$, the normalized concentration at the innermost cell  increases from $C/C_0=33$ in the neutral case to $C/C_0=101$ for $q=5 \times 10^{-15} \ \mathrm{C}$ and $C/C_0=118$ for $q=1 \times 10^{-14} \ \mathrm{C}$.

\begin{figure}
    \centering
    \includegraphics[width=13.5cm]{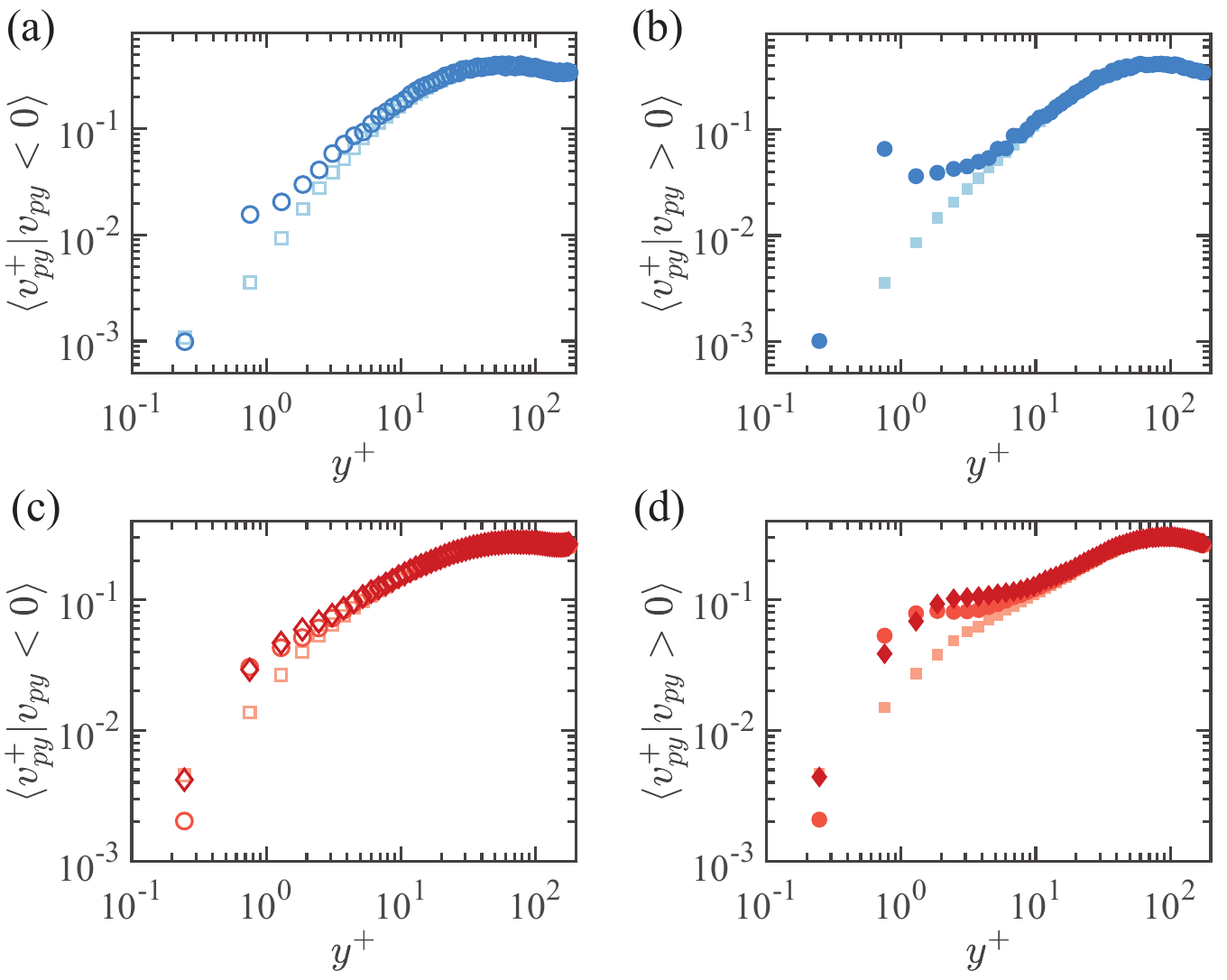}
    \caption{Normalized mean velocity for (a) approaching particles with $St^+=32$, (b) departing particles with $St^+=32$, (c) approaching particles with $St^+=133$, and (d) departing particles with $St^+=133$. Colors from light to dark represent results for $q=0 \ \mathrm{C}$, $5 \times 10^{-15} \ \mathrm{C}$, and $1 \times 10^{-14} \ \mathrm{C}$.}
    \label{fig:MeanVelocity}
\end{figure}

Apart from particle concentration, the mean approaching/departing velocity of particles is also of interest, as it describes how quickly particles located at a given $y^+$ move towards or away from the wall. Here, we define the direction pointing away from the wall as the positive direction, so the mean approaching and departing velocities normalized by $u_\tau$ are computed as $\langle v^+_{py} |v_{py} < 0 \rangle$ and $\langle v^+_{py} |v_{py} > 0 \rangle$, respectively. 

Figure \ref{fig:MeanVelocity}(a) and (c) shows the approaching velocity for $St^+=32$ and $St^+=133$. When particles are close to the channel center, the electrostatic forces pointing towards both walls cancel out, so the approaching velocity for charged particles collapses with that of neutral ones. As particles get closer to the walls, they are accelerated by the electrostatic force towards the closer wall, and the approaching velocity becomes higher than the neutral cases. The increase of the approaching velocity becomes more significant as $y^+$ decreases, indicating the more significant role played by the electrostatic force in the near-wall region. 

However, the approaching velocity at the walls (the leftmost points in figure \ref{fig:MeanVelocity} (a) and (c)) is smaller than the neutral velocity for both $St^+$. This can be attributed to several reasons. First, the two-way coupling effect caused by the concentrated particles near the wall could effectively weaken local flows, thereby reducing the wall-normal particle velocity. Although the domain-averaged particle volume fraction is as low as $\overline{\alpha} \sim O(10^{-6})$, the near-wall particle volume fraction is more than two orders of magnitude higher (figure \ref{fig:Concentration}), which is sufficient to modulate the near-wall local flows \citep{ElghobashiASR1994, BalachandarARFM2010}. Besides, after bouncing off the wall, particles must overcome the electrostatic force to become re-entrained into the outer layer. Consequently, more charged particles are trapped in the viscous layer and adjust to the low fluid velocity. Meanwhile, high-speed particles are energetic enough to escape and become re-entrained. This biased sampling of high-speed particles leads to the increase in the mean departing velocity in figure \ref{fig:MeanVelocity}(b) and (d).

With both particle concentration and wall-normal velocity, we can define the wall-normal particle flux $k$, which measures the number of particles crossing a wall-parallel plane per unit time per unit area. The wall-normal particle flux towards ($-$) and away from the wall ($+$) can be given by

\begin{subequations}
    \begin{equation}
        k^{(-)}(y)= \langle v_{py} |v_{py}<0 \rangle(y) \cdot C(y) \cdot P(v_{py}<0|y),
    \end{equation}

    \begin{equation}
        k^{(+)}(y)= \langle v_{py} |v_{py}>0 \rangle(y) \cdot C(y) \cdot P(v_{py}>0|y).
    \end{equation}
\end{subequations}

\noindent
Here, $P(v_{py}<0|y)$ and $P(v_{py}>0|y)$ are the proportions of particles moving towards and away from the walls at $y$. Note that after normalizing the particle flux $k$ as 

\begin{equation}
    k^{+,(-)}=\frac{k^{(-)}}{u_\tau C_0}, \ \text{and} \ k^{+,(+)}=\frac{k^{(+)}}{u_\tau C_0},
\end{equation}

\noindent
the dimensionless particle flux $k^+$ has the same physical meaning as the dimensionless deposition velocity defined in other works \citep{GuhaARFM2008, FongJFM2019}.

\begin{figure}
    \centering
    \includegraphics[width=13.5cm]{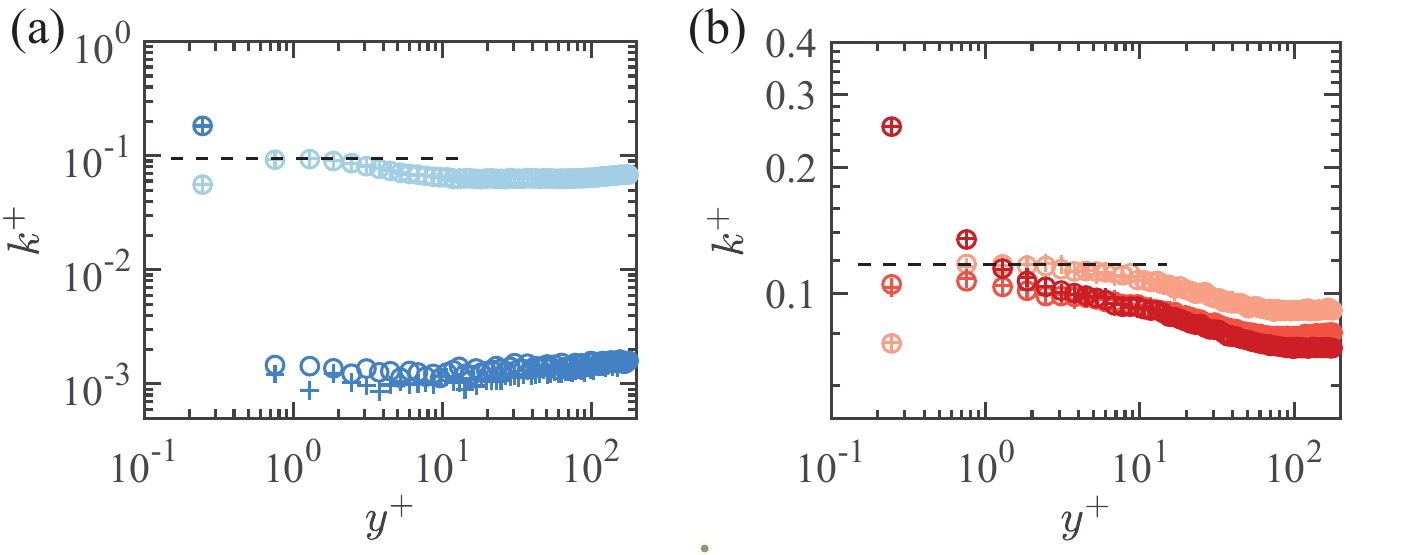}
    \caption{Dimensionless particle flux $k^+$ for (a) $St^+=32$ and (b) $St^+=133$. Circles ($\circ$) and plus signs ($+$) represent approaching and departing fluxes. The horizontal black dashed lines indicate neutral deposition velocity $k_d^+$. Colors from light to dark represent results for $q=0 \ \mathrm{C}$, $5 \times 10^{-15} \ \mathrm{C}$, and $1 \times 10^{-14} \ \mathrm{C}$.}
    \label{fig:Flux}
\end{figure}

Figure \ref{fig:Flux} displays profiles of $k^+$ for different $St^+$ and $q$. For neutral particles, one notices that the dimensionless flux $k^+$ is not constant and shows a similar trend along the $y$ direction for both $St^+=32$ and $St^+=133$. This trend is consistent with the particle transport mechanisms described in previous works \citep{SoldatiIJMF2009, ChenPRF2022}. Particles in the buffer layer ($5 \leq y^+ \leq 30$) are swept by quasi-streamwise vortices and obtain a net drift velocity towards the near-wall region, which accounts for the rise of $k^+$ in the buffer layer. Then particles trapped near the wall could either deposit at the wall after decelerating in the viscous layer, or be re-entrained to the outer layer by ejections, both of which will reduce $k^+$ in the viscous layer ($ y^+ \leq 5$). Compared with $St^+=32$, particles with $St^+=133$ are more inertial and undergo a weaker deceleration, so $k^+$ is less decreased in the viscous layer. When particles are charged, the electrostatic force becomes more dominant as particles get closer to walls, so $k$ keeps increasing as $y^+$ decreases. As most of the charged particles are concentrated near the wall (figure \ref{fig:Concentration}), a sharp increase in the near-wall flux and a decrease in the far-field flux are observed.

To make a direct comparison with the classic model prediction, it is necessary to define the deposition velocity of the particles. In previous experimental investigations, deposition velocity was obtained by directly measuring the total number (or mass) of droplets or particles deposited onto the wall in each test \citep{FriedlanderIEC1957, LiuJAS1974}. However, in this study, the process of particle sticking and deposition onto the wall is not included. Therefore, a specific wall-normal location $y^+$ must be selected, and the dimensionless deposition velocity $k^+_d$ is defined as the local particle flux $k^+(y^+)$. 

For neutral particles, the deposition velocity at $y^+=1$ is chosen (black dashed lines in figure \ref{fig:Flux}) for two reasons. First, dust particles typically have a finite size comparable to the viscous length, making the deposition velocity at $y^+ \sim 1$, just before particles bounce off, more relevant. Additionally, $k^+$ plateaus near $y^+=1$, which represents the maximum rate at which turbulence transports particles towards the wall before they are slowed down in the viscous layer. 

For charged particles, the electrostatic-enhanced accumulation occurs in the innermost cells before particles bounce off. Since the particle size used in the simulation is small, the deposition velocity at the innermost cell is chosen as $k^+_d$. If a larger particle size is used, the flux profiles of charged particles are expected to shift towards larger $y^+$, leading to a different deposition velocity. Despite these variations, the change in deposition velocity $k^+_d$ due to particle charging is expected to remain consistent.

\begin{figure}
    \centering
    \includegraphics[width=8cm]{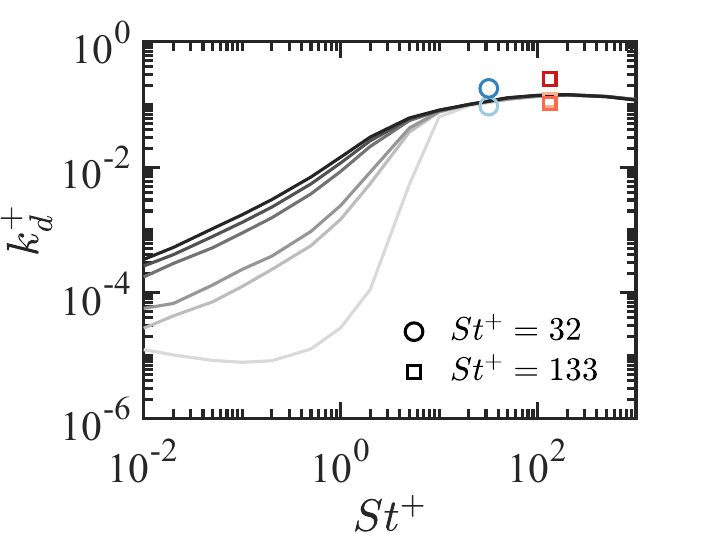}
    \caption{Comparison of deposition velocity $k^+_d$ between the current work (scatters) and the model prediction by \citet{GuhaARFM2008} (solid lines). Solid lines from light to dark are results from the charging parameter $\xi=0$, $0.05$, $0.1$, $0.5$, $0.75$, and $1$. Colors from light to dark represent results for $q=0 \ \mathrm{C}$ ($\xi=0$), $5 \times 10^{-15} \ \mathrm{C}$ ($\xi=0.16$), and $1 \times 10^{-14} \ \mathrm{C}$ ($\xi=0.31$).}
    \label{fig:DepVel_Compare}
\end{figure}

The deposition velocities $k^+_d$ from simulations are then compared with the predictions using the 1D Eulerian model by \citet{GuhaARFM2008} in figure \ref{fig:DepVel_Compare}. In the reference, the particle charge is measured by the charging parameter $\xi=q/q_{max}$ with the max particle charge $q_{max}$ depending on the particle size. Plugging in the parameters from table \ref{tab:SimulationParameters} then leads to $\xi=0.16$ for $q=5 \times 10^{-15} \ \mathrm{C}$ and $\xi=0.31$ for $q=1 \times 10^{-14} \ \mathrm{C}$. In the neutral case, the deposition velocities for both $St^+$ are close to the model prediction. However, a discrepancy arises in the charged case. While Guha's model predicts little difference in the deposition velocity of charged particles with $St^+ \geq 10$, our simulation results suggest that this may not be the case. Therefore, the physical mechanisms that enhance the deposition velocity of charged particles in the current simulations need to be further examined in the following sections.

\subsection{Driving mechanisms of wall-normal particle accumulation}
\label{sec:Model}

In the previous section, it has been demonstrated that the electrostatic force increases the deposition velocity for particles with $St^+ =32$ and $St^+=133$. While both particle concentration (figure \ref{fig:Concentration}) and wall-normal velocities (figure \ref{fig:MeanVelocity}) are affected, the change in particle concentration is more significant, which makes a predominant contribution to the increased deposition velocity. This finding also confirms that the assumption of an unchanged concentration profile for charged particles in Guha's model is invalid. In this section, we focus on the changes in particle concentration under the influence of electrostatic forces. 

To quantify the contributions of different physical mechanics to the wall-normal particle distribution, the statistical approach proposed by \citet{JohnsonJFM2020} is employed here. This one-dimensional model was originally developed for neutral particles, and has been recently extended to include charged particles  \citep{DiRenzoPRF2019, ZhangJFM2023}. For completeness, key aspects of the model are introduced in Appendix \ref{sec:Appd_Model_Derivation}, with further details available in the cited references. When the particle phase reaches equilibrium, the steady concentration profile can be given by

\begin{equation}
\label{eq:Concentration}
C(y)= \mathcal{C}^{\prime} \mathrm{exp} (I_{turb} + I_{bias} +I_{elec}).
\end{equation}

\noindent
Three different integrals are defined in the exponent of (\ref{eq:Concentration}):

\begin{subequations}
\label{eq:Integrals}
    \begin{equation}
    \label{eq:Int_a}
        I_{turb}=-\int_0^y \frac{\mathrm{d} \ln{ \langle v_{py}^2 | \eta \rangle}}{\mathrm{d} \eta} \mathrm{d} \eta,
    \end{equation}

    \begin{equation}
    \label{eq:Int_b}
        I_{bias}=\frac{1}{\tau_p} \int_0^y \frac{\langle f_I (u_{fy}-v_{py}) |\eta \rangle}{\langle v_{py}^2 |\eta \rangle} \mathrm{d} \eta,
    \end{equation}

    \begin{equation}
    \label{eq:Int_c}
        I_{elec}=\frac{q}{m} \int_0^y \frac{\langle E_{y} | \eta \rangle}{\langle v_{py}^2 | \eta \rangle} \mathrm{d} \eta.
    \end{equation}
\end{subequations}

\noindent
Here, $v_{py}$ is the wall-normal particle velocity, $u_{fy}$ is the wall-normal fluid velocity at the particle location, $f_{I}$ is the Schiller-Naumann correction factor for the drag force, and $E_y$ is the wall-normal electric field at the particle location. 

The unknown coefficient $\mathcal{C}^{\prime}$ in (\ref{eq:Concentration}) can be determined as follows. In the steady state, we first compute the mean profiles of the wall-normal particle kinetic energy $\langle v_{py}^2 \rangle(y)$, the wall-normal drag force $\langle f_I (u_{fy}-v_{py})\rangle (y)$, and the wall-norm electric field $\langle E_{y}\rangle(y)$. Then for each cell center location $y$, the integrals $I_{turb}(y)$, $I_{bias}(y)$, $I_{elec}(y)$ are obtained by integrating the corresponding terms from the innermost cell to the current cell at $y$ following \ref{eq:Integrals}. Because of particle mass conservation, the mean particle concentration $C_0$ across the channel can be related to the concentration profile $C(y)$ by 

\begin{equation}
\label{eq:Coeff}
     C_0 = \frac{1}{\delta} \int_0^{\delta} C(y) \mathrm{d}y = \frac{\mathcal{C}^{\prime}}{\delta} \int_0^{\delta} \mathrm{exp} \left( I_{turb}(y) + I_{bias}(y) +I_{elec}(y) \right) \mathrm{d}y.
\end{equation}

\noindent
In (\ref{eq:Coeff}), both $C_0$ and $\delta$ are constants. By integrating the exponential of the sum of the integrals, the unknown coefficient $\mathcal{C}^{\prime}$ can be determined.

Since knowledge of both the particle phase and the fluid phase is required, (\ref{eq:Concentration}) is not capable of predicting the steady concentration profile a priori unless additional model closures are included \citep{ZhangPRF2023}. However, these integrals provide insights into the essential roles played by different mechanisms. The first integral $I_{turb}$ depends on the gradient of the wall-normal particle kinetic energy and is referred to as the turbophoresis effect. Since $\langle v_{py}^2 \rangle$ increases as $y$ increases, $I_{turb}$ is always negative. According to (\ref{eq:Concentration}), the negative $I_{turb}$ reduces $C(y)$ as $y$ increases, which means turbophoresis drives particles towards the walls. The second integral $I_{bias}$ quantifies the slip velocity experienced by particles, and is referred to as the biased-sampling effect. In a steady state, inertial particles tend to oversample fluids moving away from the wall, leading to a positive slip velocity that pushes particles away from the walls. The last integral $I_{elec}$ depends on the electrostatic force acting on the charged particles. As will be shown below, the average wall-normal electric field points towards the wall (negative), which consistently attracts particles towards the wall, further contributing to the deposition of charged particles.

\begin{figure}
    \centering
    \includegraphics[width=13.5cm]{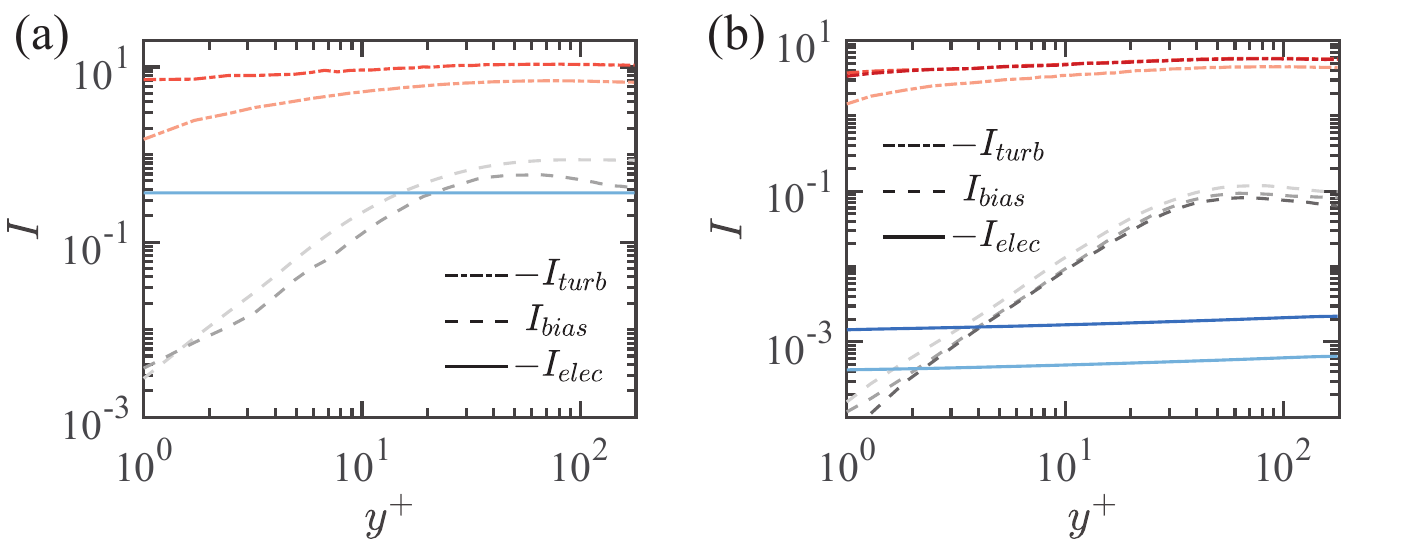}
    \caption{Comparison of different integrals for particles with (a) $St^+=32$ and (b) $St^+=133$. Colors from light to dark represent results for $q=0 \ \mathrm{C}$, $5 \times 10^{-15} \ \mathrm{C}$, and $1 \times 10^{-14} \ \mathrm{C}$.}
    \label{fig:Integrals}
\end{figure}

The wall-normal profiles of the integrals for both neutral and charged particles are displayed in figure \ref{fig:Integrals}. By taking the integrals into (\ref{eq:Integrals}) and determining the coefficient $\mathcal{C}^{\prime}$ from (\ref{eq:Coeff}), the predicted concentration profiles are plotted in Fig. \ref{fig:Concentration} as dashed lines, which show good agreement with the simulation results. (\ref{eq:Concentration}) is thus justified and the relative importance of various mechanisms can be directly quantified by comparing the values of the integrals. As shown in figure \ref{fig:Integrals}, in the neutral case, the magnitude of $I_{turb}$ is more than one order of magnitude larger than that of $I_{bias}$. Therefore, neutral particles are primarily driven by turbophoresis and exhibit significant accumulation near the wall, while the biased-sampling effect plays a secondary role in pushing particles away and counteracts the turbophoresis effect. 

When particles are charged, the electrostatic force influences particle distribution in multiple ways. First, the wall-normal electrostatic force appears in the electric integral term $I_{elec}$, which points towards the wall and directly enhances particle accumulation (\ref{eq:Int_c}). Since the magnitude of $I_{elec}$ depends on the charge-to-mass ratio ($q/m$) of particles, the direct influence of $I_{elec}$ is more important for particles with $St^+=32$ (figure \ref{fig:Integrals}(a)) than for those with $St^+=133$ (figure \ref{fig:Integrals}(b)). Furthermore, the turbophoresis term $I_{turb}$ and the biased-sampling term $I_{bias}$ are also altered for charged particles, indicating that the electrostatic force has more complex and indirect effects on particle concentration. Notably, an increase in $I_{turb}$ and a decrease in $I_{bias}$ both contribute to a higher near-wall concentration. In the following sections, we will discuss the indirect electrostatic effects through $I_{turb}$ and $I_{bias}$ and the direct electrostatic effects through $I_{elec}$.

\subsection{Turbophoresis and biased sampling of charged particles}
\label{sec:IndirectEffect}

\begin{figure}
    \centering
    \includegraphics[width=13.5cm]{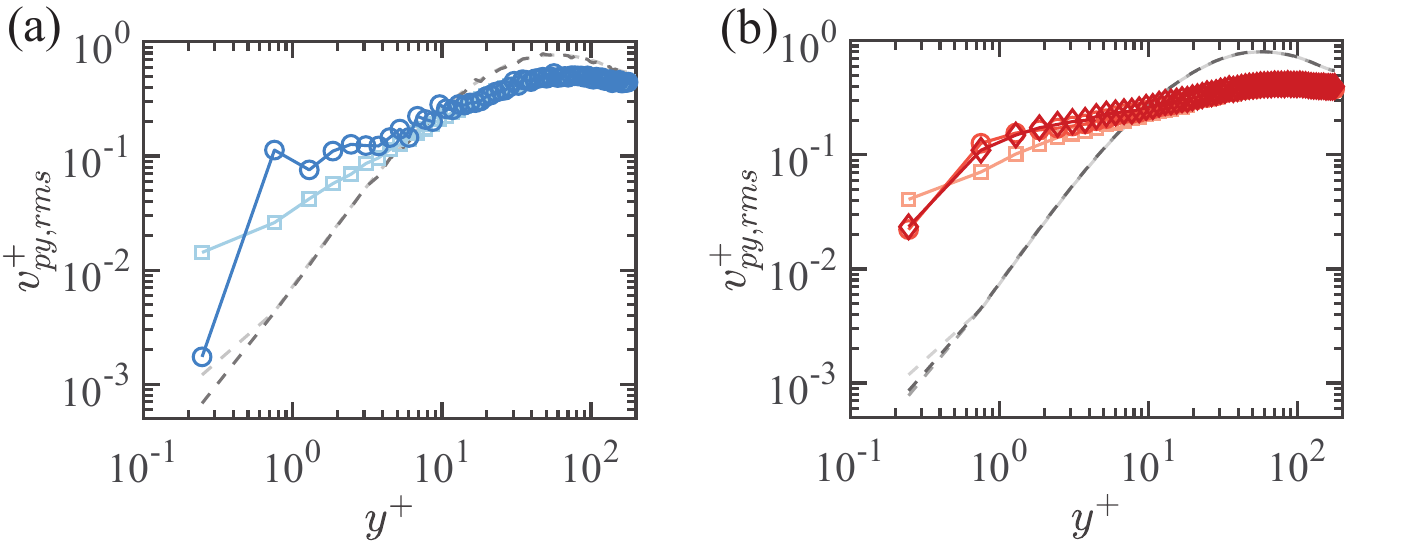}
    \caption{Dimensionless root-mean-square of wall-normal particle velocity $v^+_{py,rms}$ for (a) $St^+=32$ and (b) $St^+=133$. Dashed lines are dimensionless root-mean-square of wall-normal fluid velocity $u^+_{fy,rms}$ sampled at particle locations. Colors from light to dark represent results for $q=0 \ \mathrm{C}$, $5 \times 10^{-15} \ \mathrm{C}$, and $1 \times 10^{-14} \ \mathrm{C}$.}
    \label{fig:RMS_Velocity}
\end{figure}

To understand how the electrostatic force modulates turbophoresis, the root-mean-square (RMS) of the wall-normal particle velocity ($v_{py,rms}=\langle v_{py}^2 \rangle^{1/2}$) normalized by $u_\tau$ is presented in figure \ref{fig:RMS_Velocity}. The comparison with neutral results shows that the changes in $v_{py,rms}$ due to the electrostatic force are qualitatively similar to the changes observed in the mean wall-normal particle velocities, as seen in figure \ref{fig:MeanVelocity}. As discussed above, charged particles located outside the innermost cell exhibit higher mean wall-normal velocities because of electrostatic forces (figure \ref{fig:MeanVelocity}). This increased wall-normal velocity facilitates the transport of particles from the more energetic outer layer to the less energetic near-wall wall. Since inertial particles retain a memory of their path history, an increased RMS velocity outside the innermost cell is observed compared to the neutral results. 

In contrast, a significant drop in $v_{py,rms}^+$ is seen in the innermost cell for charged particles, creating a sharp gradient of $v_{py,rms}^+$ near the wall. This decrease can be attributed to two main reasons. (1) Two-way coupling effect: The high particle concentration near the wall reduces the local turbulent intensity, leading to a corresponding decrease in particle kinetic energy. This is evidenced by the decrease in the fluid RMS velocity $u^+_{fy,rms}$ sampled at particle locations and shown in figure \ref{fig:RMS_Velocity} as dashed lines. (2) Longer residence time in the viscous layer: Charged particles trapped in the viscous layer require more energetic ejections to overcome the electrostatic attraction and be re-entrained into the outer layer. This leads to a longer residence time in the viscous layer. Consequently, charged particles interact with the near-wall low-speed fluid for a longer period and their RMS velocity is effectively damped. For particles with $St^+=32$, the RMS velocity becomes one order of magnitude smaller than that of neutral particles, while for particles with $St^+=133$ that tend to retain their original RMS velocity for a longer period, $v_{py,rms}^+$ is still reduced by half. This change can also be understood from an energy perspective: due to electric potential energy, particles transfer turbulent kinetic energy from the outer layer to the near-wall region, where it is eventually dissipated through fluid drag.

Such a non-trivial change in the RMS velocity profile can significantly influence the turbophoresis effect. As expressed by (\ref{eq:Int_a}), $I_{turb}$ depends on the relative change of the wall-normal kinetic energy:

\begin{align*}
    \frac{\mathrm{d} \ln{ \langle v_{py}^2 \rangle}}{\mathrm{d} y}=\frac{1}{\langle v_{py}^2 \rangle} \frac{\mathrm{d} \langle v_{py}^2 \rangle}{\mathrm{d} y}.
\end{align*}

\noindent
Thus, the reduced RMS velocity close to the wall and the enhanced RMS velocity slightly away from the wall leads to a sharp gradient of RMS velocity near the wall and a significant increase in $I_{turb}$ as shown in figure \ref{fig:Integrals}. In contrast, the 1D Eulerian model did not account for the complex changes in particle RMS velocity profile due to the electrostatic force and the particle-fluid coupling effects. It instead relies on the local fluid properties to relate the unknown particle RMS velocity profile to the prescribed fluid RMS velocity profile \citep{GuhaARFM2008}. As a result, the 1D Eulerian model is unable to predict the modulation of turbophoresis. It is important to emphasize that, this substantial rise in $I_{turb}$ is the primary factor behind the increased concentration of charged particles at the wall, which in turn results in a higher deposition velocity.

We now turn to how the electrostatic force modulates the biased-sampling effect. The biased-sampling effect is closely related to the interaction between inertial particles and the near-wall coherent structures. Therefore, we employ the quadrant analysis to quantify how particles sample different fluid structures in the buffer layer. In this analysis, the fluctuations of the streamwise and the wall-normal fluid velocities sampled at the particle locations are denoted by $u_{fx}^{\prime}$ and $u_{fy}^{\prime}$. Four quadrants can be defined based on the signs of $u_{fx}^{\prime}$ and $u_{fy}^{\prime}$. In particular, ejection events correspond to outward motion of low-speed fluid ($u_{fx}^{\prime}<0$, $u_{fy}^{\prime}>0$), while sweep events (Q4) correspond to inward motion of high-speed fluid ($u_{fx}^{\prime}>0$, $u_{fy}^{\prime}<0$).

\begin{figure}
    \centering
    \includegraphics[width=13.5cm]{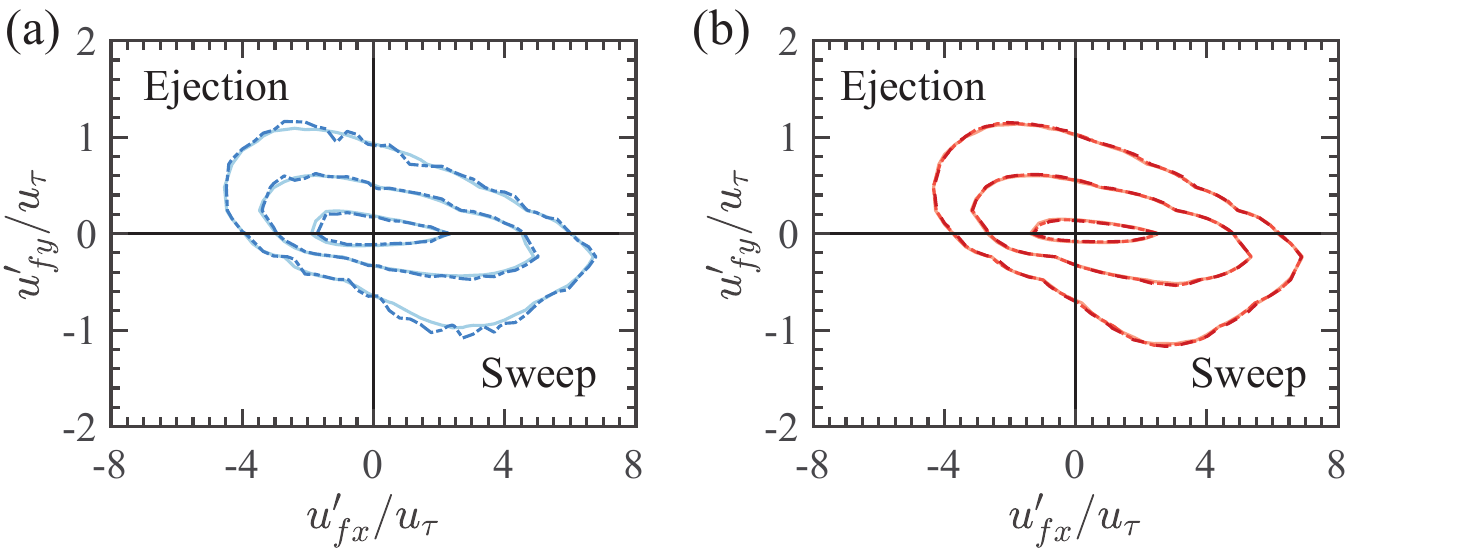}
    \caption{Joint PDF of the streamwise and the wall-normal fluid velocity fluctuations, $u_{fx}^{\prime}$ and $u_{fy}^{\prime}$, at the particle locations for (a) $St^+=32$ and (b) $St^+=133$ within the range $5 \leq y^+ \leq 30$. Contours from inside out represent a value of $0.01$, $0.05$, and $0.2$, respectively. Colors from light to dark represent results for $q=0 \ \mathrm{C}$, $5 \times 10^{-15} \ \mathrm{C}$, and $1 \times 10^{-14} \ \mathrm{C}$.}
    \label{fig:Quadrant}
\end{figure}

Figure \ref{fig:Quadrant} shows the joint probability density functions (PDF) of the particle-sampled fluid velocity fluctuations at $5 \leq y^+ \leq 30$. For both $St^+=32$ and $St^+=133$, neutral particles show a tendency to sample Q2 and Q4 more frequently than Q1 and Q3. This confirms that ejections (Q2) and sweeps (Q4) play a dominant role in transporting particles near the wall. In figure \ref{fig:Quadrant}(a), contours of the joint PDF of charged particles ($St^+=32$, $q=5 \times 10^{-15} \ \mathrm{C}$) are less smooth because of the lower particle concentration within the range $5 \leq y^+ \leq 30$ (figure \ref{fig:Concentration}(a)). Despite the reduced particle concentration, the general shape of the contours in the charged case remains similar to those of neutral particles. However, two differences are also observed: the portion of particles in Q2 decreases, while the portion of particles in Q4 increases. This trend is better highlighted by comparing the proportions of particles sampling Q2 and Q4, as summarized in table \ref{tab:Quadrant}. As a result, charged particles sample less upward fluid velocities than neutral particles, which explains the consistent decrease of $I_{bias}$ with the increase of $q$ for $St^+=32$ in figure \ref{fig:Integrals}(a). The same trend is also observed for $St^+=133$ particles. However, since $St^+=133$ particles are more inertial, the change in figure \ref{fig:Quadrant}(b) is less significant, leading to a smaller change in $I_{bias}$ (figure \ref{fig:Integrals} (b)).  

\begin{table}
  \begin{center}
\def~{\hphantom{0}}
  {\begin{tabular}{l|cc|cc}
     & \multicolumn{2}{c|}{$St^+=32$} & \multicolumn{2}{c}{$St^+=133$}\\
     & $Q2$ & $Q4$ & $Q2$ & $Q4$\\
     \
    $q=0 \ \mathrm{C}$ & 35.73 \% & 32.59\% & 29.57\% & 38.49 \% \\
    $q=5 \times 10^{-15} \ \mathrm{C}$ & 33.95 \% & 34.47\% & 29.36\% & 38.79 \% \\
    $q=1 \times 10^{-14} \ \mathrm{C}$ & - & - & 29.23\% & 38.90 \% \\
  \end{tabular}}
  \caption{Proportion of particles sampling Q2 and Q4 within the range $5 \leq y^+ \leq 30$.}
  \label{tab:Quadrant}
  \end{center}
\end{table}

It is noteworthy that the observed trend of sampling less upward flows is similar to the phenomenon of preferential sweeping in the gravitational settling of heavy particles in turbulence. It is known that heavy particles settling in HIT may tend to sample the downward-velocity region of vortices, aligning with the direction of gravity. This behavior leads to an enhanced average settling velocity of inertial particles \citep{WangJFM1993, BecPRL2014}. Similarly, in wall-bounded turbulence, an analogous enhancement in settling velocity has been reported. Particles subject to a constant force directed towards the wall preferentially sample flow regions that are also moving towards the wall as they pass through the buffer layer, which effectively increases particles' settling velocity \citep{ChenPRF2022}. Given that fluid sweeps are typically more intense and spatially concentrated than ejections, the bias introduced by the electrostatic force is even stronger compared to that in HIT. In the current study, the electrostatic force acting on charged particles plays a similar role to gravity in these previous studies. The particles are driven towards the wall by the electrostatic attraction, leading them to preferentially sample fluid motions that also move towards the wall. This leads to a reduction in upward-flow sampling (ejections) and an increase in downward-flow sampling (sweeps), thereby making a secondary contribution to the accumulation of particles near the wall. Furthermore, the electrostatic force is not uniform across the entire channel but becomes stronger closer to the wall, making its influence on biased sampling an increasingly important factor to consider.

In addition, in HIT, the gravitational settling velocity of heavy particles can be either enhanced or reduced, depending on the ratio of gravitational settling velocity to turbulence intensity. Accordingly, we expect to observe different regimes based on the relative importance of the wall-pointing electrostatic force compared to turbulent fluctuations. In this study, due to the low particle charge and concentration, the particle-induced electric field remains weak. As a result, the wall-pointing electrical migration velocity is small relative to turbulent fluctuations, which lies within the regime of preferential sweeping that leads to enhanced deposition. However, if the electrical migration velocity becomes more significant, such as in the presence of a strong external field or with highly charged particles, this enhancement may change. For instance, when the electrical migration velocity greatly exceeds turbulent fluctuations, particle behavior may decouple from near-wall coherent structures, and the deposition enhancement is suppressed. However, when the electrical migration velocity becomes comparable to turbulent fluctuations, it remains unclear whether particles will experience a slowdown due to the loitering effect, as reported in previous works in HIT. This presents an interesting topic for future investigations.

\subsection{Wall-normal electric field}
\label{sec:Efield}

In this section, we discuss the profile of the wall-normal electric field $E_y$, which directly affects particle concentration through $I_{elec}$. Moreover, $E_y$ also serves as a direct indicator of the significance of the electrostatic force on particle behavior. 

As suggested in \citet{GuhaARFM2008}, a particle $i$ with charge $q_i$ near a grounded conducting wall experiences the electrostatic force due to the induced charge on the wall, which equals the Coulomb force from its image located at the symmetric location about the wall with the opposite charge $-q_i$. If the particle-wall distance is $y_w$, the wall-normal electric field due to the PW interaction can be computed by

\begin{equation}
\label{eq:PW_EField}
    E_y^{(Im)}=-\frac{q_i}{4 \pi \varepsilon_0 (2 y_{w,i} )^2}=-\frac{q_i}{16 \pi \varepsilon_0 y_{w,i}^2}.
\end{equation}

\begin{figure}
    \centering
    \includegraphics[width=13.5cm]{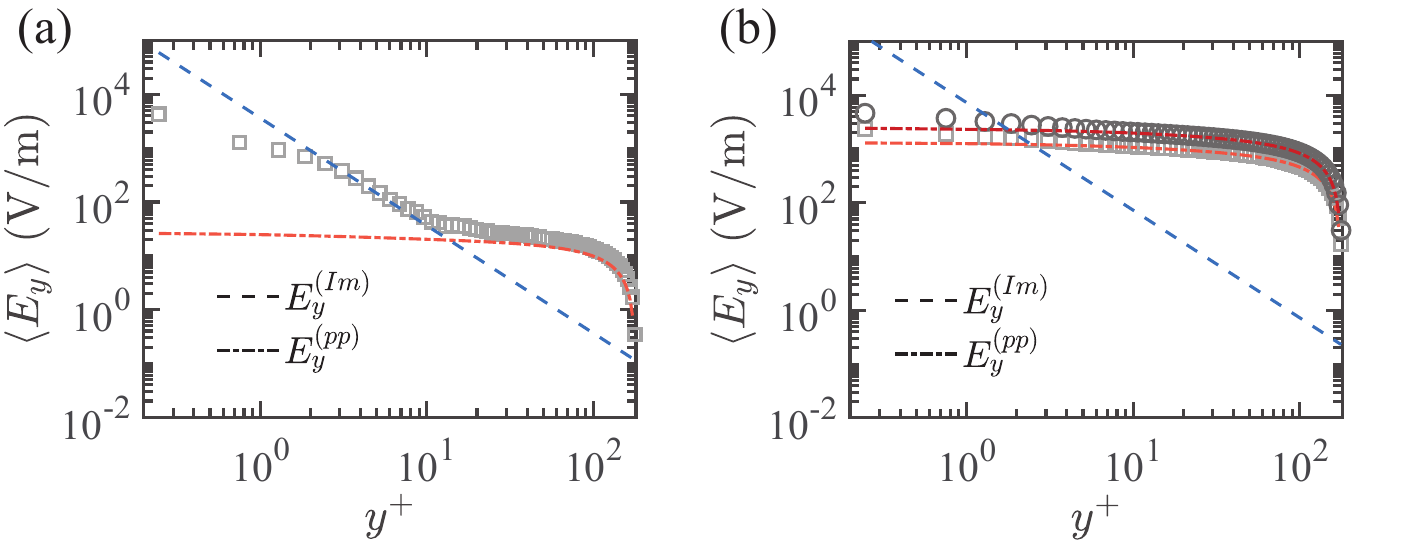}
    \caption{Averaged wall-normal electric field $\langle E_y \rangle$ for particles with (a) $St^{+}=32$ and (b) $St^{+}=133$. Scatters with light to dark grey correspond to a particle charge of $q=1 \ \times 10^{-15} \ \mathrm{C}$, and \ $2 \ \times 10^{-15} \ \mathrm{C}$. Contributions from the particle-wall (PW) and particle-particle (PP) electrostatic interactions are shown as blue dashed lines and red dash-dotted lines, respectively.}
    \label{fig:ElectricField}
\end{figure}

The average wall-normal electric field $\langle E_y \rangle$ for particles with $St^+=32$ and $q=5 \times 10^{-15} \ \mathrm{C}$ is then compared with $E_y^{(Im)}$ in figure \ref{fig:ElectricField}(a). One notices that $E_y^{(Im)}$ collapses with the simulation results only within the intermediate range of $ 2 \leq y^+ \leq 10$, while significant deviations occur in both the near-wall and far-field regions. These deviations indicate that the PW interaction alone cannot account for all the electrostatic forces acting on particles, highlighting the need to include the electric field generated by the particle-particle (PP) electrostatic interaction. Thus, we derive the electric field due to the PP interaction, $\mathbf{E}^{(pp)}$, starting from Gauss law

\begin{equation}
\label{eq:GaussLaw_1}
    \nabla \cdot \mathbf{E}^{(pp)}=\frac{\rho_c}{\varepsilon_0} = \frac{qC}{\varepsilon_0}.
\end{equation}

\noindent
Here, the volumetric charging density $\rho_c$ equals the product of the particle charge $q$ and the particle concentration $C$. Taking the ensemble average of (\ref{eq:GaussLaw_1}) leads to 

\begin{equation}
\label{eq:GaussLaw_2}
    \frac{\mathrm{d} \langle E_y^{(pp)} \rangle}{\mathrm{d} y}=\frac{q\langle C \rangle}{\varepsilon_0}.
\end{equation}

\noindent
Note that in (\ref{eq:GaussLaw_2}) the electric field components in periodic directions become zero after ensemble averaging, i.e., $\langle E_x^{(pp)} \rangle=\langle E_z^{(pp)} \rangle=0$. Integrate (\ref{eq:GaussLaw_2}) from a certain location $y$ to the centerline $\delta$ then yields

\begin{align*}
    \langle E_y^{(pp)} \rangle(\delta)-\langle E_y^{(pp)} \rangle(y)=\frac{q}{\varepsilon_0}\int_{\eta=y}^{\eta=\delta} \langle C \rangle(\eta) \mathrm{d}\eta.
\end{align*}

\noindent
Considering the symmetry of the system, the wall-normal electric field at the centerline is zero, i.e., $\langle E_y^{(pp)} \rangle(\delta)=0$. Therefore, the PP electric field is

\begin{equation}
\label{eq:PP_Efield}
    \langle E_y^{(pp)} \rangle(y)=-\frac{q}{\varepsilon_0}\int_{\xi=y}^{\eta=\delta} \langle C \rangle(\eta) \mathrm{d}\eta.
\end{equation}

In (\ref{eq:PP_Efield}), $\langle E_y^{(pp)} \rangle(y)$ is negative, indicating the PP electrostatic force always points towards the wall. Specifically, for a target particle located at $y$, the net PP electrostatic force equals the Coulomb repulsion from all the particles located between $y$ to the centerline $\delta$, which pushes the target particle towards the wall. The PP electric field $\langle E_y^{(pp)} \rangle(y)$ is then plotted in figure \ref{fig:ElectricField}(a) as a red dash-dotted line, which agrees with the simulation results in the far-field region ($y^+ \geq 20$). 

We therefore propose three distinct regions of the wall-normal electric field, as illustrated in figure \ref{fig:ElectricField}(a). In the far-field region at large $y^+$, the contribution from the PW interaction is negligible compared to the PP interaction. Particles in this region are primarily driven towards the wall by PP Coulomb repulsion. As $y^+$ decreases, $\langle E_y^{(Im)} \rangle(y)$ levels off as the integral in (\ref{eq:PP_Efield}) saturates, while the PW interaction continues to rise and eventually becomes dominant. Consequently, the PW interaction prevails as the primary electrostatic force in the intermediate region. Finally, when the particle approaches the wall, the repulsion from the concentrated particles counteracts the PW attraction, resulting in $\langle E_y \rangle(y)$ being lower than that predicted by (\ref{eq:PW_EField}).

Interestingly, not all three regions exist in all cases, as shown in figure \ref{fig:ElectricField}(b) for charged particles with $St^+=133$. The transition between the intermediate and the far-field regions depends on the relative importance of the PW and the PP interactions:

\begin{equation}
\label{eq:EFieldModels}
     \underbrace{\frac{q}{16 \pi \varepsilon_0 y_{w}^2}}_\text{PW interaction}, \   \text{and} \ \underbrace{\frac{1}{\varepsilon_0}\int_{\eta=y}^{\eta=\delta} [q \times \langle C \rangle(\eta)] \mathrm{d}\eta}_\text{PP interaction}.
\end{equation}
   
\noindent
As shown in figure \ref{fig:Concentration}(b), the particle concentration $C(y)$ for $St^+=133$ is high in the outer flow, leading to a more pronounced PP interaction. Consequently, the PP interaction dominates nearly up to the wall. In such cases, relying solely on the image charge force would significantly underestimate the magnitude of the electrostatic force. 

In the end, discussing the relative importance of the PW and the PP electrostatic interactions across a broader range of scenarios is essential for developing a more complete understanding of the electrostatic effects arising from particle charging. In this study, the particles are monodispersed and identically charged, meaning that the net charge between $y$ and $\delta$ in (\ref{eq:PP_Efield}) is always non-zero, resulting in a net repulsive force. The significance of this repulsion depends on the net charge distribution within the channel. With a much lower particle concentration, the PP interaction is expected to be less influential, allowing the PW interaction to dominate at larger $y^+$. In addition, for monodispersed particles carrying both positive and negative charges, as is common in triboelectrification, the PP interaction becomes negligible because the integral of the net charge in (\ref{eq:PP_Efield}) equals zero. However, in more complex systems with bidispersed oppositely charged particles, the concentration profiles for different particle groups will differ. Even if the overall system is neutral, there will be a separation between the centers of positive and negative centers. Consequently, the PP interaction will migrate light particles accumulated near the wall outward, while attracting heavy particles dispersed in the outer layer towards the wall, as reported by \citet{ZhangJFM2023}. Finally, beyond the channel flow investigated here, the transport of charged particles in turbulent boundary layers, such as sandstorms and pollutants dispersion in the atmosphere, is also widespread. In these systems, where there is only one wall, the PP interaction can still be evaluated by adjusting the upper limit:

\begin{align*}
    \langle E_y^{(pp)} \rangle(y)=\langle E_y^{(pp)} \rangle(y_{Ref})-\frac{1}{\varepsilon_0}\int_{\eta=y}^{\eta=y_{Ref}} q\langle C \rangle(\eta) \mathrm{d}\eta.
\end{align*}

\noindent
Here, $\langle E_y^{(pp)} \rangle(y_{Ref})$ is the electric field at a reference point $y_{Ref}$. Thus, the PP interaction may still play a role as long as the net charge integral is significant.

\section{Conclusions}
This work utilizes four-way coupled simulations to address an important question: how particle charging affects the deposition velocity of particles onto an electrically grounded conductor through a turbulence boundary layer, particularly in the context of charged particle deposition in gas turbines. In this study, we developed a canonical case involving charged particles transported in a fully developed turbulent channel flow. Contrary to previous model predictions, which suggested no change in deposition velocity when particles are inertial and dominated by the turbophoresis effect, we found that electrostatic forces actually increase the deposition velocity.

Since the increase in the deposition velocity of charged particles primarily results from the enhanced near-wall accumulation, the wall-normal profile of charged particles is further examined. By employing a statistical approach in the particle phase space $(y,v_{py})$, three mechanisms affecting the concentration profiles can be quantified in the form of integrals: turbophoresis ($I_{turb}$), biased sampling ($I_{bias}$), and electrostatic forces ($I_{elec}$). It was found that the electrostatic force creates a sharper gradient in the wall-normal particle RMS velocity, which significantly increases $I_{turb}$. As a result, the enhanced turbophoresis effect is identified as the main driver of the more extreme particle accumulation near the wall. In addition, charged particles are found to sample upward flow regions less frequently than neutral particles, which reduces the biased-sampling effect $I_{bias}$. This change occurs because charged particles subject to the wall-pointing electrostatic force tend to sample the downward-moving fluids as they pass through coherent structures in the buffer layer. This behavior is analogous to the preferential sweeping effect observed in the settling of heavy particles in turbulence. Finally, the profile of the wall-normal electric field is discussed. It is found that both the particle-wall (PW) interaction and the particle-particle (PP) interaction contribute to the electrostatic force acting on charged particles. Depending on the conditions, the relative importance of the PW and PP interactions results in distinct electric field profiles. Consequently, when the net charge carried by suspended particles is significant, relying solely on the classic image charge model may lead to a significant underestimation of the electrostatic effects. 

According to the original framework of \citet{GuhaARFM2008}, the deposition velocity incorporates contributions from both the wall-normal particle concentration and velocity. To predict the deposition velocity for charged particles, it is assumed (1) that the particle velocity is modulated solely by the image force, and (2) that the particle concentration remains unchanged. Upon carefully analyzing our simulation results, these assumptions are found to be invalid. First, the wall-normal electrostatic force comprises contributions from both particle-particle (PP) and particle-wall (PW) interactions, whereas the classic model only accounts for the latter. In certain cases, such as figure \ref{fig:ElectricField}(b), this omission leads to a significant underestimation of the magnitude of the electrostatic force. Second, as shown in figure \ref{fig:Concentration}, electrostatic forces drastically modulate particle concentration, which is the primary contributor to the increased deposition in this study. This critical effect is entirely absent in the classic model. Given that the 1D Eulerian model has been widely used across various communities, it is crucial to highlight these limitations to ensure proper interpretation and application.

Regarding the physical process itself, several findings about its highly coupled nature are also presented. First, and most counterintuitively, the influence of electrostatic force is affected by particle-turbulence interaction. Since the PP electrostatic force depends on the concentration profile (\ref{eq:PP_Efield}), the spatial distribution of particles determines the dominant electrostatic force, as illustrated by the distinct electric field profiles shown in figure \ref{fig:ElectricField}. Consequently, a careful comparison of the relative importance of PP and PW electrostatic interactions is necessary. In contrast, many earlier studies often assumed the dominance of the image force without question. Second, turbophoresis, the primary mechanism that shapes the particle profile (figure \ref{fig:Integrals}), is found to be highly sensitive to the wall-normal RMS velocity of the particles ($v_{p,rms}^+$). Even a subtle change in $v_{p,rms}^+$ (figure \ref{fig:RMS_Velocity}) can lead to a drastic change in the particle concentration profile. Therefore, in future studies, any factor that might affect $v_{p,rms}^+$ should be treated carefully, such as electrostatic forces, two-way coupling, and particle-particle collisions. Moreover, although the current system is dilute, the effects of two-way coupling and interparticle collisions should still be accounted for, as the nonuniform particle concentration may locally transition into the two-way or four-way coupled regime.

Finally, it would also be valuable to discuss the potential influences of various parameters, such as the turbulence Reynolds number and particle inertia, on the findings of this study. The motivation for this work is to investigate the dust ingestion problem in jet engines. Due to the small characteristic length scales of the internal flow and the higher fluid viscosity at elevated operating temperatures \citet{LawsonJT2011}, the friction Reynolds number is not expected to be extremely high. For example, the diameter of the cooling hole is given as $1.69\times 10^{-3} \ \mathrm{m}$ in \citet{LawsonJT2011} and $4.6\times 10^{-3} \ \mathrm{m}$ in \citet{LawsonTurboExpo2010}. By choosing the radius of the hole as the half channel width $\delta$, and considering the friction velocity $u_{\tau}=3.59 \ \mathrm{m/s}$, the fluid density $\rho_f=3.32 \ \mathrm{kg/m^3}$ and the fluid viscosity $\mu_f=5.55 \times 10^{-5} \ \mathrm{Pa \cdot s}$ in Section 2.1, the friction Reynolds number lies within the range of 180 to 490. Thus, the chosen $Re_{\tau}$ is within the parameter space for internal deposition. For external deposition, the Reynolds number may be even higher because of the high speed and large length scales. However, the key physics that drives particle deposition, i.e., particle inertia and electrostatics, will remain valid. Therefore, we choose $Re_{\tau} = 180$ to keep the flow configurations similar to those in our experimental investigations on the transport and deposition of charged inertial particles in a vertical turbulent channel, where $Re_{\tau} \approx 200$. As discussed by \citet{JohnsonJFM2020}, the transport mechanisms (turbophoresis and biased sampling) of neutral particles appear consistent across multiple Reynolds numbers. Consequently, the modulation of particle deposition velocity by electrostatic forces is also expected to remain consistent, allowing the findings of this study to be extended to high Reynolds numbers that are more representative of realistic flow conditions.

Although inertial particles are discussed in this study, how electrostatic force affects the deposition of tracer-like particles is also relevant in many applications. For inertialess particles, the contributions of turbophoresis and biased sampling are no longer present, meaning that the enhancement of particle deposition arises only from the direct effect of $I_{elec}$. In this case, the wall-normal electrostatic force becomes the primary mechanism that enhances deposition, which depends on the particle charging conditions. In a system where tracer particles carry both positive and negative charges, the particle-particle (PP) electrostatic force ($E_y^{(pp)}$) is zero, and the dominant force is the image force ($E_y^{(Im)}$) due to the particle-wall (PW) electrostatic force. According to \citet{YaoPT2021}, under these conditions, tracers follow local fluid motions faithfully when away from the wall, but detach from the local flow and accelerate towards the wall as they approach the near-wall region, where the image force becomes significant. As a result, the influence of PW interaction is limited to the near-wall region. If tracers are identically charged, in addition to the image forces, the PP electrostatic repulsion ($E_y^{(pp)}$) contributes significantly to the far field (figure \ref{fig:ElectricField}). Consequently, tracer trajectories may detach from local streamlines even when they are still far from the wall. Meanwhile, as the electrostatic force grows increasingly significant near the wall, the particle slip velocity will show a continuous increase as particles approach the wall. In addition to the driving mechanism, the resistance to tracer deposition is also of interest. For inertial particles, biased sampling serves as the primary mechanism that pushes particles away from the wall. However, this mechanism is absent for tracers. Consequently, the resistance to tracer deposition is also expected to arise from the electrostatic force. For identically charged tracers, as particles accumulate near the wall, the mutual repulsion between them also grows. If the wall is not grounded or is made of dielectric material, the local electric potential continues to rise, which effectively repels new incoming particles. As a result, a balance is established between the wall-approaching attraction and the near-wall repulsion, leading to a steady state. However, if (i) the particles carry opposite charges, eliminating mutual repulsion, or (ii) the wall is conducting and grounded, causing the mutual repulsion to be largely suppressed by the image force effect, the electric potential near the wall will remain close to zero. Consequently, there is no resisting force to prevent particle deposition. In this case, a steady state cannot be achieved, and all particles will eventually migrate towards the wall and become captured.
\\ 

\section*{Acknowledgements}
This work was supported by the Office of Naval Research (ONR) under Grant NO. N00014-21-1-2620. This work was also partially supported by an Early Stage Innovation grant from NASA's Space Technology Research Grants Program under Grant NO. 80NSSC21K0222.

\section*{Declaration of interests}
The authors report no conflict of interest.

\appendix

\section{Assessment of grid resolution}
\label{sec:Appd_GridResolution}
In the main text, a Cartesian grid with a resolution of $128^3$ is used. The grid is uniform in both the $x$ and $z$ directions, and stretched in the $y$ direction with a stretching factor of $S=1.9$. To assess grid sensitivity, we simulated test cases on a refined grid with a resolution of $256^3$ and the same stretching factor. The grid information is summarized in table \ref{tab:GridInfo}.

\begin{figure}
    \centering
    \includegraphics[width=10cm]{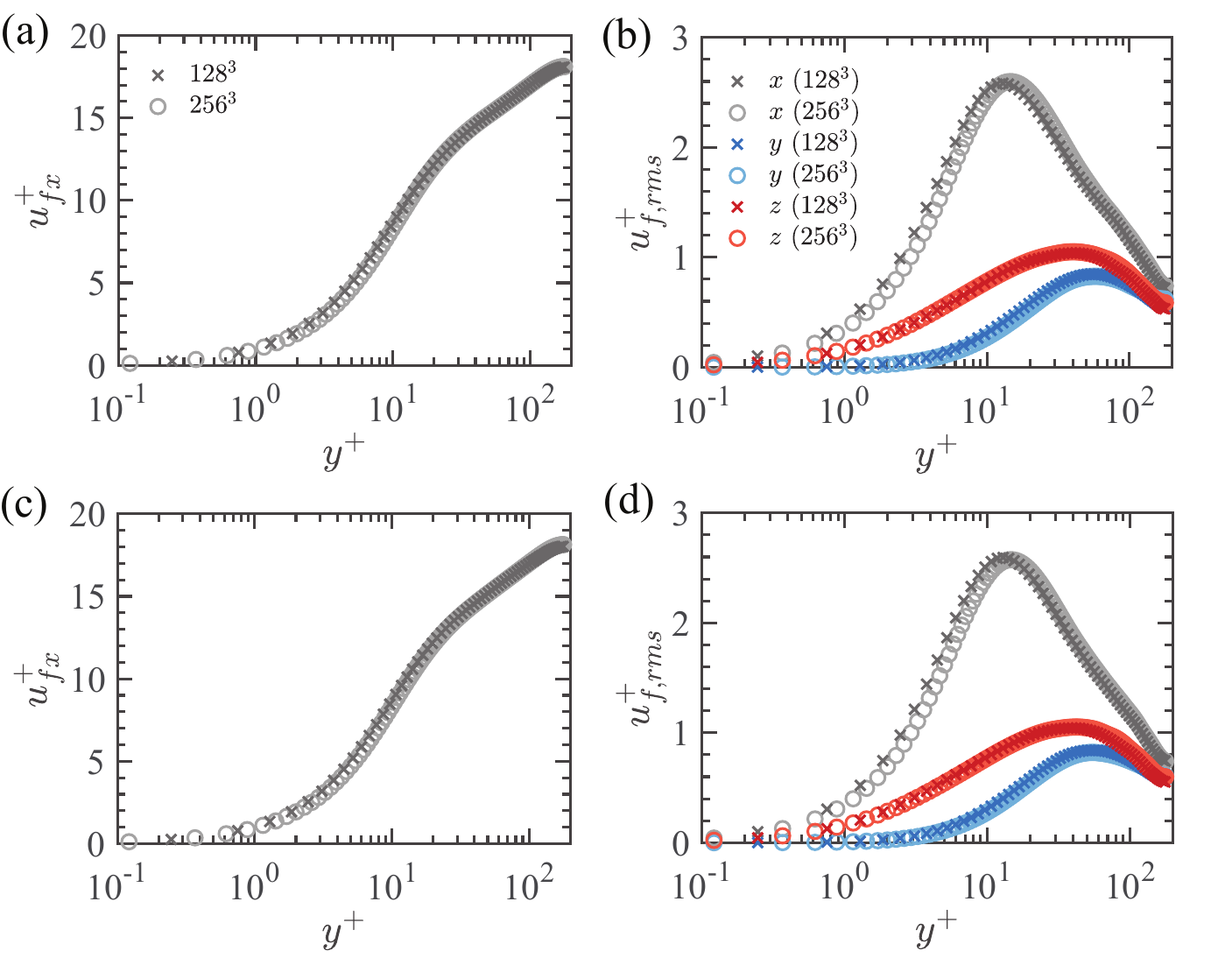}
    \caption{Mean streamwise fluid velocity in the case with (a) $St^+=32$ and (c) $St^+=133$. Root-mean-square of fluid fluctuation velocity in $x$, $y$, $z$ directions for (b) $St^+=32$ and (d) $St^+=133$. Crosses ($x$) represent results using the original grid mesh ($128^3$), and circles ($\circ$) denote results using a refined mesh ($256^3$).}
    \label{fig:Grid_Fluid}
\end{figure}

\begin{figure}
    \centering
    \includegraphics[width=13.5cm]{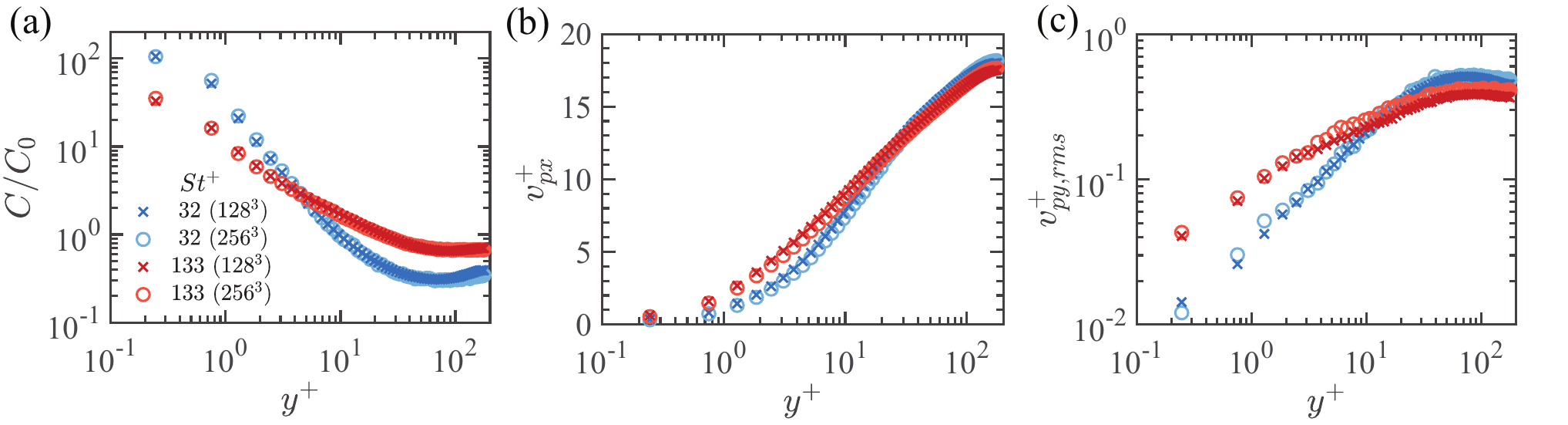}
    \caption{Comparison of (a) normalized wall-normal particle concentration $C/C_0$, (b) mean streamwise particle velocity, and (c) root-mean-square of wall-normal particle fluctuation velocity. Crosses ($x$) represent results using the original grid mesh ($128^3$), and circles ($\circ$) denote results using a refined mesh ($256^3$).}
    \label{fig:Grid_Particle}
\end{figure}

Two different Stokes numbers ($St^+=32/133$) are used in the tests while the particle charge is set to zero. The fluid velocity profiles for the two grid resolutions are shown in figure \ref{fig:Grid_Fluid}, while particle concentration and velocity profiles are compared in figure \ref{fig:Grid_Particle}. Most fluid and particle statistics remain unchanged when the mesh is refined. The wall-normal particle RMS velocity (figure \ref{fig:Grid_Particle}(c)) shows a slight increase near the channel center on the refined mesh. However, this does not lead to any significant modulation in particle concentration, as observed in figure \ref{fig:Grid_Particle}(a). Therefore, the grid resolution of $128^3$ used in the main text is deemed sufficient. \\

\begin{table}
  \begin{center}
  \begin{tabular}{lllllll}
    Grid & Resolution & $L_x \times L_y \times L_z$ & $S$  & $\Delta x^+$ & $\Delta y^+$ & $\Delta z^+$\\
    Original & $128^3$ & $4 \pi \times 2 \times 2 \pi$ & $1.9$ & $17.67$ & $0.49-5.58$ & $8.84$\\
    Refined & $256^3$ & $4 \pi \times 2 \times 2 \pi$ & $1.9$ & $8.84$ & $0.24-2.78$ & $4.42$\\
  \end{tabular}
  \caption{Summary of grid assessment.}
  \label{tab:GridInfo}
  \end{center}
\end{table}

\section{Volume-filtered Eulerian-Lagrangian framework}
\label{sec:Appendix_VFEL}
This section presents a brief derivation of the governing equations of the volume-filtered Eulerian-Lagrangian (VFEL) framework employed in this work. During the derivation, certain simplifications are made to obtain the final form presented in the main text. Justifications for these simplifications are also provided below. Further details about the VFEL framework can be found in \citet{CapecelatroJCP2013, AndersonIECF1967}. 

\subsection{Governing equations of fluid motion}
In the standard point-wise Eulerian-Lagrangian approach, the governing equations of the fluid phase without body forces are

\begin{subequations}
\begin{equation}
\label{eq:pointContinuity}
    \frac{\partial{\rho_f}}{\partial t} + \nabla \cdot (\rho_f \mathbf{u}_f) = 0, 
\end{equation}

\begin{equation}
\label{eq:pointMomentum}
    \frac{\partial (\rho_f \mathbf{u}_f)}{\partial t} + \nabla \cdot (\rho_f \mathbf{u}_f \otimes \mathbf{u}_f) = \nabla \cdot \mathbf{\tau}.
\end{equation}
\end{subequations}

\noindent
Here $\rho_f$ and $\mathbf{u}_f$ are the density and velocity of the fluid. The fluid stress is given by

\begin{equation}
    \mathbf{\tau}=-p \mathbf{I} + \mu_f \left[ (\nabla \mathbf{u}_f+{\nabla \mathbf{u}_f}^T) - \cfrac{2}{3}(\nabla \cdot \mathbf{u}_f)\mathbf{I} \right],
\end{equation}

\noindent
where $p$ and $\mu_f$ are the fluid pressure and dynamic viscosity. A Gaussian filter $G_F$ is then defined as

\begin{equation*}
    G_F(r)=\frac{1}{\sqrt{2 \pi}\sigma} \exp{\left( -\frac{r^2}{2 \sigma^2} \right)}.
\end{equation*}

\noindent
The filter length $\delta_F$, defined as the width of $G_F(r)$ at the half height, can be related to $\sigma$ as $\delta_F=2 \sqrt{2 \ln{2}} \sigma$. The fluid volume fraction can then be defined as

\begin{equation}
    \alpha_f(\mathbf{x},t)=\int_{\mathcal{V}_f} G_F(|\mathbf{x}-\mathbf{y}|) \mathrm{d} \mathbf{y},
\end{equation}

\noindent
where $\mathcal{V}_f$ means the integral is taken over all points $\mathbf{y}$ occupied by the fluid phase. Applying the Gaussian filter to any point property $\mathbf{a}(\mathbf{x},t)$ of the fluid then yields

\begin{equation}
    \alpha_f \overline{\mathbf{a}}(\mathbf{x},t) = \int_{\mathcal{V}_f} \mathbf{a}(\mathbf{x},t) G_F(|\mathbf{x}-\mathbf{y}|) \mathrm{d} \mathbf{y},
\end{equation}

\noindent
where $\overline{\mathbf{a}}(\mathbf{x},t)$ is the volume-filtered property. The associated residual can be written as $\mathbf{a}^{\prime} (\mathbf{x},t)=\mathbf{a}(\mathbf{x},t)-\overline{\mathbf{a}}(\mathbf{x},t)$.

We now derive the volume-filtered motion equations. By assuming that the shortest distance from $\mathbf{x}$ to the boundaries of the system is much larger than the filter size, \citet{AndersonIECF1967} derived the volume filtering of the temporal derivative, divergence and gradient of a point property as 

\begin{subequations}
    \begin{equation}
    \label{eq:VF_dt}
        \int_{\mathcal{V}_f} \frac{\partial \mathbf{a}(\mathbf{y},t)}{\partial t} G_F(|\mathbf{x}-\mathbf{y}|) \mathrm{d} \mathbf{y}= \frac{\partial}{\partial t}(\alpha_f \overline{\mathbf{a}}(\mathbf{x},t))+\sum_{i=1}^{N_p}\int_{\mathcal{S}_i} \mathbf{n} \cdot \mathbf{u}_i \mathbf{a}(\mathbf{y},t)G_F(|\mathbf{x}-\mathbf{y}|)\mathrm{d} \mathbf{y},
    \end{equation}

    \begin{equation}
    \label{eq:VF_divergence}
        \int_{\mathcal{V}_f} \nabla \cdot \mathbf{a}(\mathbf{y},t) G_F(|\mathbf{x}-\mathbf{y}|) \mathrm{d} \mathbf{y}=\nabla \cdot (\alpha_f \overline{\mathbf{a}}(\mathbf{x},t))-\sum_{i=1}^{N_p}\int_{\mathcal{S}_i} \mathbf{n}\cdot\mathbf{a}(\mathbf{y},t)G_F(|\mathbf{x}-\mathbf{y}|)\mathrm{d} \mathbf{y},
    \end{equation}

    \begin{equation}
    \label{eq:VF_gradient}
        \int_{\mathcal{V}_f} \nabla \mathbf{a}(\mathbf{y},t) G_F(|\mathbf{x}-\mathbf{y}|) \mathrm{d} \mathbf{y}=\nabla (\alpha_f \overline{\mathbf{a}}(\mathbf{x},t))-\sum_{i=1}^{N_p}\int_{\mathcal{S}_i} \mathbf{n}\otimes\mathbf{a}(\mathbf{y},t)G_F(|\mathbf{x}-\mathbf{y}|)\mathrm{d} \mathbf{y}.
    \end{equation}

\end{subequations}

\noindent
Here, $\mathcal{S}_i$ represents the spherical surface of particle $i$. $\mathbf{n}$ is the outward unit vector on the particle surface, and $\mathbf{u}_i$ denotes the velocity of the solid matter at point $\mathbf{y}$ on $\mathcal{S}_i$. Since there is no mass transfer between the solid and fluid phases, $\mathbf{u}_i$ is equal to the fluid velocity at the particle surface. 

For a constant-density fluid, multiplying (\ref{eq:pointContinuity}) by $G_F$ and integrating over $\mathcal{V}_f$, followed by the application of (\ref{eq:VF_dt}) and (\ref{eq:VF_divergence}), yields the volume-filtered continuity equation

\begin{equation}
\label{eq:VF_Continuity}
    \frac{\partial \alpha_f}{\partial t} + \nabla \cdot (\alpha_f \overline{\mathbf{u}}_f)=0.
\end{equation}

\noindent
Similarly, volume filtering the left-hand side of (\ref{eq:pointMomentum}) and again applying (\ref{eq:VF_dt}) and (\ref{eq:VF_divergence}) result in

\begin{equation}
\label{eq:VF_Momentum_LHS}
    \frac{\partial}{\partial t}(\alpha_f \rho_f \overline{\mathbf{u}}_f) + \nabla \cdot (\alpha_f \rho_f \overline{\mathbf{u}}_f \otimes \overline{\mathbf{u}}_f) + \nabla \cdot (\alpha_f \rho_f \overline{\mathbf{u}_f^\prime \otimes \mathbf{u}_f^\prime}),
\end{equation}

\noindent
where the residual Reynolds stress is
\begin{equation}
    \mathcal{F}_{u}=\nabla \cdot (\alpha_f \rho_f \overline{\mathbf{u}_f^\prime \otimes \mathbf{u}_f^\prime}).
\end{equation}

The volume filtering of the right-hand side of (\ref{eq:pointMomentum}) can be obtained by substituting $\mathbf{a}=\mathbf{\tau}$ in (\ref{eq:VF_divergence}), which reads

\begin{equation}
\label{eq:VF_RHS}
    \int_{\mathcal{V}_f} \nabla \cdot \mathbf{\tau}(\mathbf{y},t) G_F(|\mathbf{x}-\mathbf{y}|) \mathrm{d} \mathbf{y}=\nabla \cdot (\alpha_f \overline{\mathbf{\tau}})-\sum_{i=1}^{N_p}\int_{\mathcal{S}_i} \mathbf{n}\cdot\mathbf{\tau}(\mathbf{y},t)G_F(|\mathbf{x}-\mathbf{y}|)\mathrm{d} \mathbf{y},
\end{equation}

\noindent
where the filtered stress is written as 
\begin{equation}
\label{eq:VF_Stress}
    \overline{\mathbf{\tau}}=\overline{\mathbf{\tau}}^* + \mathbf{\tau}_{\mu} = -\overline{p} \mathbf{I} + \mu_f \left[ (\nabla \overline{\mathbf{\mathbf{u}}}_f+{\nabla \overline{\mathbf{u}}_f}^T) - \cfrac{2}{3}(\nabla \cdot \overline{\mathbf{u}}_f)\mathbf{I} \right] + \mathbf{\tau}_{\mu}.
\end{equation}

\noindent
Here, $\overline{\mathbf{\tau}}^*$ is the nominal stress evaluated using the filtered velocity field $\overline{\mathbf{u}}_f$. The residual stress $\mathbf{\tau}_{\mu}$ is defined as the differences between $\overline{\mathbf{\tau}}$ and $\overline{\mathbf{\tau}}^*$:

\begin{equation}
\label{eq:tau_mu}
    \mathbf{\tau}_{\mu} = \mu_f \left[ (\overline{\nabla \mathbf{\mathbf{u}}_f}+\overline{\nabla \mathbf{u}_f^T}) - (\nabla \overline{\mathbf{\mathbf{u}}}_f+{\nabla \overline{\mathbf{u}}_f}^T) - \cfrac{2}{3}(\overline{\nabla \cdot \mathbf{u}_f} - \nabla \cdot \overline{\mathbf{u}}_f)\mathbf{I} \right].
\end{equation}

\noindent
Note that if the fluid dynamic viscosity is modulated by the particle phase, as is typical in dense particulate flows, the modulation of $\mu_f$ would introduce an additional contribution to $\mathbf{\tau}_{\mu}$. However, in this study, the particle phase is dilute, so $\mu_f$ is treated as constant. 

The second term on the right-hand side of (\ref{eq:VF_RHS}) can be decomposed into the contributions from the volume-filtered stress ($\overline{\mathbf{\tau}}$) and the residual stress ($\mathbf{\tau}^\prime$). Because the filter size is large compared to the particle diameter ($\delta_F=8d_p$), the filtered stress $\overline{\mathbf{\tau}}$ varies little at the particle scale, so it can be taken out of the integral. As a result, the contribution from the volume-filtered stress can be simplified as

\begin{equation*}
    \sum_{i=1}^{N_p}\int_{\mathcal{S}_i} \mathbf{n}\cdot \overline{\mathbf{\tau}}(\mathbf{y},t)G_F(|\mathbf{x}-\mathbf{y}|)\mathrm{d} \mathbf{y} \approx \overline{\mathbf{\tau}} \cdot \nabla \alpha_f.
\end{equation*}

\noindent
The right-hand side of (\ref{eq:VF_RHS}) can then be reorganized as

\begin{align}
    & \nabla \cdot (\alpha_f \overline{\mathbf{\tau}}) - \overline{\mathbf{\tau}} \cdot \nabla \alpha_f - \sum_{i=1}^{N_p}\int_{\mathcal{S}_i} \mathbf{n}\cdot\mathbf{\tau}^\prime(\mathbf{y},t)G_F(|\mathbf{x}-\mathbf{y}|)\mathrm{d} \mathbf{y} \\
    = & \nabla \cdot \overline{\mathbf{\tau}} -\left( \alpha_p  \nabla \cdot \overline{\mathbf{\tau}} + \sum_{i=1}^{N_p}\int_{\mathcal{S}_i} \mathbf{n}\cdot\mathbf{\tau}^\prime(\mathbf{y},t)G_F(|\mathbf{x}-\mathbf{y}|)\mathrm{d} \mathbf{y} \right),
    \label{eq:VF_Couple1}
\end{align}

\noindent
where $\alpha_p=1-\alpha_f$ is the filtered particle volume fraction. We now show that, the last two terms in (\ref{eq:VF_Couple1}) are related to the interphase force coupling. For an individual particle $i$, the fluid force can be simplified as

\begin{equation}
\label{eq:FluidForce}
    \mathbf{F}_i^F=\int_{\mathcal{S}_i}\mathbf{\tau}\cdot \mathbf{n} \mathrm{d}\mathbf{y}=\int_{\mathcal{S}_i} (\overline{\mathbf{\tau}} + \mathbf{\tau}^\prime) \cdot \mathbf{n} \mathrm{d}\mathbf{y} = \int_{\mathcal{V}_i} \nabla \cdot \overline{\mathbf{\tau}} \mathrm{d}\mathbf{y} + \int_{\mathcal{S}_i} \mathbf{\tau}^\prime \cdot \mathbf{n} \mathrm{d}\mathbf{y} \approx \nabla \cdot \overline{\mathbf{\tau}} \mathcal{V}_i +\int_{\mathcal{S}_i} \mathbf{\tau}^\prime \cdot \mathbf{n} \mathrm{d}\mathbf{y}.
\end{equation}

\noindent
As the particle size is much smaller than the filter size, $\overline{\mathbf{\tau}}$ varies little at the particle scale and is again taken out of the integral in the last step in (\ref{eq:FluidForce}). $\mathcal{V}_i$ is the volume of particle $i$. The momentum transfer term can be obtained by summing the filtered fluid force over all particles

\begin{equation}
    \mathcal{F}_P= \sum_{i=1}^{N_p} G_F(|\mathbf{x}-\mathbf{y}|) \mathbf{F}_i^F = \alpha_p  \nabla \cdot \overline{\mathbf{\tau}} + \sum_{i=1}^{N_p}\int_{\mathcal{S}_i} \mathbf{n}\cdot\mathbf{\tau}^\prime(\mathbf{y},t)G_F(|\mathbf{x}-\mathbf{y}|)\mathrm{d} \mathbf{y}.
    \label{eq:VF_Couple2}
\end{equation}

Finally, combining (\ref{eq:VF_Momentum_LHS}), (\ref{eq:VF_RHS}), (\ref{eq:VF_Stress}), (\ref{eq:VF_Couple1}) and (\ref{eq:VF_Couple2}) yields the volume-filtered fluid momentum equation

\begin{equation}
\label{eq:VF_Momentum}
    \frac{\partial}{\partial t}(\alpha_f \rho_f \overline{\mathbf{u}}_f) + \nabla \cdot (\alpha_f \rho_f \overline{\mathbf{u}}_f \otimes \overline{\mathbf{u}}_f)= \nabla \cdot \overline{\mathbf{\tau}}^* +\nabla \cdot \mathbf{\tau}_{\mu} - \mathcal{F}_{u} -\mathcal{F}_P.
\end{equation}

\subsection{Model closure}
The closures of two terms, i.e., $\mathcal{F}_u$ and $\mathbf{\tau}_{\mu}$, are required. We first evaluate the importance of the residual Reynolds stress $\mathcal{F}_u$ in this study. The Reynolds stress is usually closed using a turbulent viscosity model
\begin{equation}
    \mathcal{F}_u = \rho_f \nu_t (\nabla \overline{\mathbf{u}}_f + \overline{\mathbf{u}}_f^T).
\end{equation}

\noindent
Here, the turbulent eddy viscosity is written as
\begin{equation}
\label{eq:EddyViscosity}
    \nu_t=2 {(C_S \Delta)}^2 |\overline{S}|,
\end{equation}

\noindent
where $C_S$ is the Smagorinsky coefficient, $\Delta$ is the filter width. The strain rate of the filtered fluid velocity is $\overline{S}_{ij}=(\partial \overline{u}_i/\partial x_j + \partial \overline{u}_j/\partial x_i)/2$, and $\overline{S}={(2\overline{S}_{ij} \overline{S}_{ij})}^{1/2}$. A dynamic subgrid model \citep{GermanoPOF1991, LillyPOF1992} is employed to estimate the value of $\nu_t$, and the Smagorinsky coefficient can be determined as

\begin{equation}
\label{eq:SmagorinskyCoeff}
    C_{S}^2=\frac{L_{ij}M_{ij}}{2 M_{pq}M_{pq}},
\end{equation}

\noindent
where 
\begin{subequations}
    \begin{equation}
        L_{ij}=-\widehat{\overline{u}_i \overline{u}_j} + \widehat{\overline{u}}_i \widehat{\overline{u}}_j.
    \end{equation}

    \begin{equation}
        M_{ij}={(2\delta_F)}^2 |\widehat{\overline{S}}| \widehat{\overline{S}}_{ij} - \delta_F^2 \widehat{|\overline{S}| \overline{S}_{ij}}.
    \end{equation}
\end{subequations}

\noindent
Here, the properties filtered by the Gaussian filter with a filter length $\delta_F$ are denoted by an overline ($\overline{\cdot}$). A second coarser filter with a filter length of $2\delta_F$ is then defined and the associated filtered properties are shown with a hat ($\widehat{\cdot}$). 

\begin{figure}
    \centering
    \includegraphics[width=6.5cm]{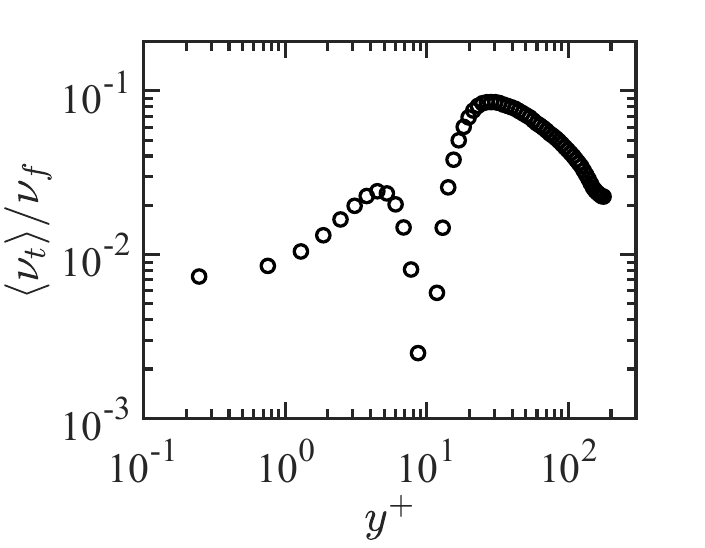}
    \caption{Ratio of plane-averaged eddy viscosity $\langle \nu_t \rangle$ to the molecular viscosity $\nu_f$ in the case with $St^+=32$ and $q=0 \ \mathrm{C}$.}
    \label{fig:EddyViscosity}
\end{figure}

Figure \ref{fig:EddyViscosity} shows the ratio of the mean eddy viscosity, $\langle \nu_t \rangle$, to the molecular viscosity, $\nu_f$, along the wall-normal direction. The mean eddy viscosity is computed by averaging $\nu_t (\mathbf{x},t)$ over the wall-parallel ($x$-$z$) plane and time. Profiles of $\langle \nu_t \rangle / \nu_f$ for other cases are not shown, as they show no noticeable differences compared to figure \ref{fig:EddyViscosity}. The ratio remains significantly smaller than unity throughout the channel, with a peak value of $ \langle \nu_t \rangle_\mathrm{max} / \nu_f = 8.5 \times 10^{-2}$. This suggests that the unresolved Reynolds stress, $\mathcal{F}_{u}$, is negligible compared to the resolved stress, $\overline{\mathbf{\tau}}^*$, and is thus omitted in this study. 

The insignificant Reynolds stress can be attributed to the fact that the filter size, $\delta_F$, is small compared to the size of near-wall coherent structures. In this study, there is a significant scale separation between the particle size ($d_p=0.36 \delta_\nu$) and the turbulent coherent structures. For example, the core size of quasi-streamwise vortices in the $x$-$z$ plane is typically around $O(10) \delta_\nu$ and even longer in the streamwise direction \citep{MarchioliJFM2002}. Consequently, even though the filter size is relatively large compared to the particle diameter ($\delta_F=8d_p$), the near-wall coherent structures remain resolved. 

We now address the closure of $\mathbf{\tau}_{\mu}$. As expressed in (\ref{eq:tau_mu}), $\mathbf{\tau}_{\mu}$ arises from the filtering of the velocity gradient. In previous studies, $\mathbf{\tau}_{\mu}$ is often modeled by introducing an effective viscosity $\mu^*$, which depends on the particle volume fraction, as seen in both dilute and dense particulate flows \citep{ZhangIJMF1997, PatankarIJMF2001}. To leading order, the relative change in viscosity scales with $\alpha_p$. In this work, given the low particle concentration, the change of $\mu^*$ is expected to be small. As a result, $\mathbf{\tau}_{\mu}$ is also omitted. Finally, by omitting both $\mathcal{F}_u$ and $\mathbf{\tau}_{\mu}$, (\ref{eq:VF_Momentum}) becomes

\begin{equation}
\label{eq:VF_Momentum_Simplified}
    \frac{\partial}{\partial t}(\alpha_f \overline{\mathbf{u}}_f) + \nabla \cdot (\alpha_f \overline{\mathbf{u}}_f \otimes \overline{\mathbf{u}}_f)= \frac{1}{\rho_f} (\nabla \cdot \overline{\mathbf{\tau}}^* -\mathcal{F}_P).
\end{equation}

\noindent
(\ref{eq:VF_Continuity}) and (\ref{eq:VF_Momentum_Simplified}) are in fact equivalent to (\ref{eq:NS equation}) without the forcing term. For simplicity, the symbols representing volume filtering are omitted in the main text.

\section{Importance of the lift force and the lubrication force}
\label{sec:Appd_FluidForce}
In this study, both the lift force and the lubrication force are omitted due to their negligible impacts under the given simulation conditions. The reasons and justifications are provided below.

\subsection{Lift force}
The extended expression of Saffman lift force is used to evaluate the importance of lift force in this study. The magnitude of the lift force is 
\begin{equation*}
    F_l=\frac{9J}{\pi} \mu_f (d_p/2)^2 u_{slip} (G/\nu_f)^{1/2}.
\end{equation*}
\noindent
Here $J$ is a coefficient to be determined, $\mu_f$ and $\nu_f$ are the dynamic and kinematic viscosity of the fluid, $d_p$ is the particle diameter, $u_{slip}$ is the particle slip velocity in the streamwise direction, and $G$ is the fluid shear rate. The magnitude of drag force can be written as $F_d=3 \pi \mu_f d_p v_{slip}$, where $v_{slip}$ is the particle slip velocity in the wall-normal direction . The ratio between the force magnitudes can then be written as
\begin{equation*}
    \frac{F_l}{F_d}=\frac{3J}{4\pi^2}Re_G^{1/2} \frac{u_{slip}}{v_{slip}},
\end{equation*}
\noindent
where $Re_G=G d_p^2/\nu_f$ is the particle shear Reynolds number. In this study, the fluid shear rate is estimated using the inner scales as $G=1/\tau_{\nu}=u_\tau^2/\nu_f$, which leads to $Re_G=0.13$. The particle Reynolds number is calculated as $Re_p=v_{slip}d_p/\nu_f \leq 1$. Based on the values of both $Re_p$ and $Re_G$, the coefficient $J$ is determined to be $J \leq 2.172$  using the fitting equation proposed by \citet{Mei1992IJMF}. Moreover, the velocity ratio $u_{slip}/v_{slip}$ ranges approximately from 2 to 5 in the simulations. Finally, the force ratio is computed as $F_l/F_d \leq 0.118-0.295$, which suggests that the influence of the lift force is minor compared to the drag force. We therefore neglect the lift force in this study.\\

\subsection{Short-range lubrication force}
In this study, the lubrication force is negligible because of the large particle-to-air density ratio ($\rho_p/\rho_f \sim O(10^3)$). In other multiphase flow systems, such as bubble flows ($\rho_p/\rho_f \sim O(10^{-3})$) or colloidal systems ($\rho_p/\rho_f \sim O(1)$), the lubrication force will be substantial and must be considered.

To confirm this argument, the influence of lubrication force can be estimated as follows. For a pair of particles ($i$ and $j$) approaching each other, the lubrication force has been derived by \citet{MarshallPOF2011}, which is given as

\begin{equation*}
    F_{lub}=\frac{3 \pi \mu_f r_p^2}{2h} \left( -\frac{\mathrm{d}h}{\mathrm{d}t} \right),
\end{equation*}

\noindent
where $h=|\mathbf{x}_i-\mathbf{x}_j|-(d_{p,i}+d_{p,j})/2$ is the gap between the surfaces of the two particles, and $(-\mathrm{d}h/\mathrm{d}t)$ is the radial approaching velocity. As two particles approach each other, they need to squeeze out the fluid film in between in order to collide, and the associated energy barrier is 
\begin{equation*}
    E_{lub}=\int_{h_{min}}^{h_{max}} F_{lub} \mathrm{d}h. 
\end{equation*}

\noindent
Here, $h_{max}$ is the initial separation distance below which the short-range lubrication effect becomes important, and $h_{min}$ represents the minimum separation distance between colliding particles. According to \citet{BarnockyJFM1989}, the fluid density and viscosity within the contact region can increase substantially due to the high pressure in the gap, exhibiting solid-like behavior and thereby imposing a lower limit on $h_{min}$. In addition, surface roughness further constrains $h_{min}$ due to the presence of finite-size asperities on the particle surfaces. Here, we set $h_{max}=0.01 r_p$ and $h_{min}= 5 \times 10^{-5} r_p$ with $r_p$ being the particle radius, which yields collision outcomes that show reasonable agreement with experimental data \citep{YangPOF2006, MarshallPOF2011}. By taking the initial approaching velocity $v_{init} \ge |\mathrm{d}h/\mathrm{d}t|$ out of the integral, the upper limit of the energy barrier becomes 

\begin{equation*}
    E_{lub}=\frac{3}{2} \pi \mu_f r_p^2 v_{init} \ln{\left( \frac{h_{max}}{h_{min}} \right)}. 
\end{equation*}

\noindent
Meanwhile, the driving force of an interparticle collision is the relative kinetic energy $E_{kin}=M v_{init}^2/2$, where $M=m/2$ is the effective mass of a two-particle system. Finally, the significance of lubrication is quantified by the energy ratio
\begin{equation*}
    \frac{E_{lub}}{E_{kin}} = \frac{9 \mu_f}{\rho_p d_p v_{init}} \ln{\left( \frac{h_{max}}{h_{min}} \right)}.
\end{equation*}

\noindent
In our simulations, $v_{init}$ is calculated as the mean radial relative velocity between a pair of approaching particles with a gap $h \in [0.009r_p ,0.011r_p]$. The resulting values are $v_{init}/u_{\tau}=0.268$ for $St^+=32$ and $v_{init}/u_{\tau}=3.382$ for $St^+=133$. Plugging in the simulation parameters then yields $E_{lub}/E_{kin}=0.110$ for $St^+=32$ and $E_{lub}/E_{kin}=0.002$ for $St^+=133$. The small energy ratios suggest that the lubrication force has a weak effect on interparticle motions during collisions. Therefore, the lubrication force is omitted in the simulations.\\

\section{Influence of interpolation scheme}
\label{sec:Appd_Interpolation}
\begin{figure}
    \centering
    \includegraphics[width=13.5cm]{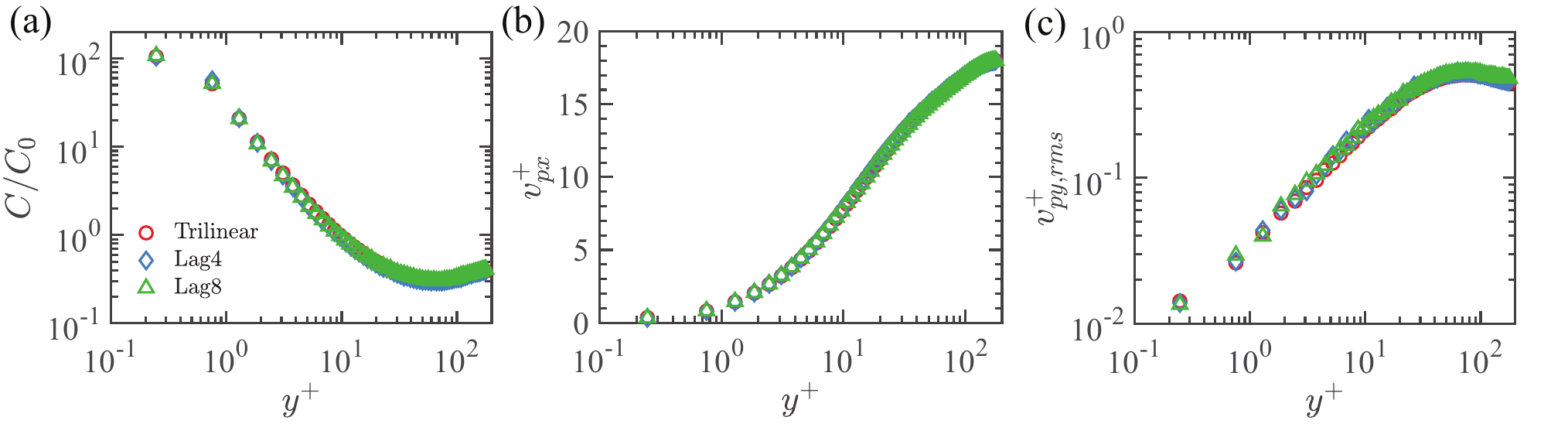}
    \caption{Comparison of (a) normalized wall-normal particle concentration $C/C_0$, (b) mean streamwise particle velocity, and (c) root-mean-square of wall-normal particle fluctuation velocity using different interpolation schemes for the case with $St^+=32$ and $q=0 \ \mathrm{C}$.}
    \label{fig:InterpolationScheme}
\end{figure}

The order of the interpolation scheme can indeed influence the accuracy of high-order derivatives of velocity. However, in this study, we only consider the drag force, which does not require higher-order derivatives of velocity at the particle position. Since the calculation of drag force depends solely on the interpolation of fluid velocity, the second-order trilinear interpolation is sufficient. To check the effect of interpolation order, the same test case ($St^+=32$, $q=0 \ \mathrm{C}$) was run using three different interpolation schemes: second-order trilinear interpolation (Trilinear), 4th-order Lagrangian interpolation (Lag4), and 8th-order Lagrangian interpolation (Lag8). Comparisons of the steady-state statistics are presented in figure \ref{fig:InterpolationScheme}. As no significant differences were observed among the results obtained with different interpolation schemes, the accuracy of trilinear interpolation is considered adequate for this study.\\

\section{Undisturbed fluid velocity in drag force calculation}
\label{sec:Appd_Undisturbed}
In this section, we discuss the error caused by self-induced disturbance in drag force in two-way coupled simulations. By definition, the Stokes drag on a target particle at $x_p$ is evaluated based on the slip velocity $(\Tilde{u}_f(x_p)-v_p)$, where $\Tilde{u}_f(x_p)$ is the undisturbed fluid velocity at the particle location. However, in two-way coupled simulations, the feedback force from the target particle itself perturbs surrounding fluid flow. As a result, the local fluid velocity, $u_f(x_p) (\neq \Tilde{u}_f(x_p))$, is disturbed, leading to an underestimated slip velocity and, consequently, a reduced drag force.

The error in drag force depends on the ratio of the particle size to the length scale of the projection scheme used to map particle disturbances onto the grid mesh. In standard Eulerian-Lagrangian simulations, where particle feedback forces are typically distributed to nearby grid points, the error scales as $O(d_p/\Delta x)$, where $\Delta x$ is the grid spacing \citep{HorwitzJCP2016}. In this study, however, we distribute the particle volume fraction and the feedback force using a Gaussian filter $G_F(r)$ with a filter length $\delta_F=8 d_p$ ((\ref{eq:GaussFilter1})) and (\ref{eq:GaussFilter2})). As the projection length scale becomes no smaller than $\delta_F$, the upper limit of the error is expected to depend on the new size ratio $d_p/\delta_f$.

Regarding the volume-filtered Eulerian-Lagrangian framework used in the present study, \citet{IrelandJCP2017} discussed the corrections of both fluid volume fraction ($\zeta_{\alpha_f}=\Tilde{\alpha}_f- \alpha_f$) and fluid velocity ($\zeta_{u_f}=\Tilde{u}_f-u_f$). By considering the case of the steady Stokes flow around a particle, the corrections can be given by:

\begin{subequations}
\begin{equation}
    \zeta_{\alpha_f}=\Tilde{\alpha}_f-\alpha_f=\mathrm{erf}\left(\frac{1}{\sqrt{2} \hat{\sigma}_c}\right)-\frac{\sqrt{2/\pi}}{\hat{\sigma}_c}\mathrm{exp}\left( -\frac{1}{2 \hat{\sigma}_c^2 }\right),
\end{equation}

\begin{equation}
    \frac{\zeta_{u_f}}{U}=\frac{\left(\frac{1}{\sqrt{2} \hat{\sigma}_c}\right)\mathrm{exp}\left( -\frac{1}{2 \hat{\sigma}_c^2 }\right)}{1-\mathrm{erf}\left(\frac{1}{\sqrt{2} \hat{\sigma}_c}\right) + \frac{\sqrt{2/\pi}}{\hat{\sigma}_c}\mathrm{exp}\left( -\frac{1}{2 \hat{\sigma}_c^2 }\right)}.
\end{equation}
\end{subequations}

\noindent
Here $\hat{\sigma}_c=(\delta_f/d_p)/\sqrt{2\ln{2}}$, and $U$ is the slip velocity in the Stokes flow problem. We would like to note that, in cases with high particle volume fraction, the drag force model typically accounts for the local fluid volume fraction ($\alpha_f=1-\alpha_p$). Consequently, the self-induced disturbance of fluid volume fraction ($\zeta_{\alpha_f}$) could also influence the drag force in two-way coupled simulations. However, in the present work, the mean particle volume fraction is low ($\overline{\alpha}_p \sim 10^{-6}$), so the drag force model does not include corrections for the influence of $\alpha_f$. Furthermore, the particle Reynolds number in the current study satisfies $Re_p \leq 1$. As a result, although the velocity correction is derived based on Stokes flow, it provides a reasonable first-order approximation of the velocity correction. Using the filter length $\delta_f=8 d_p$ then yields $\zeta_{\alpha_f}=8.42 \times 10^{-3}$ and $\zeta_{u_f}/U=1.16 \times 10^{-1}$. This indicates that the error in the drag force due to self-induced velocity disturbance is secondary. To further verify this statement, we apply the fluid velocity correction scheme proposed in \citet{IrelandJCP2017} with $\hat{\zeta}_{u_f}=\zeta_{u_f}/U=1.16 \times 10^{-1}$, which reads

\begin{equation}
    \Tilde{\mathbf{u}}_f=\frac{\mathbf{u}_f-\mathbf{v}_p \hat{\zeta}_{u_f}}{1-\hat{\zeta}_{u_f}},
\end{equation}

\noindent
to two test cases with different Stokes numbers ($St^+=32/133$) and zero particle charge.

\begin{figure}
    \centering
    \includegraphics[width=13.5cm]{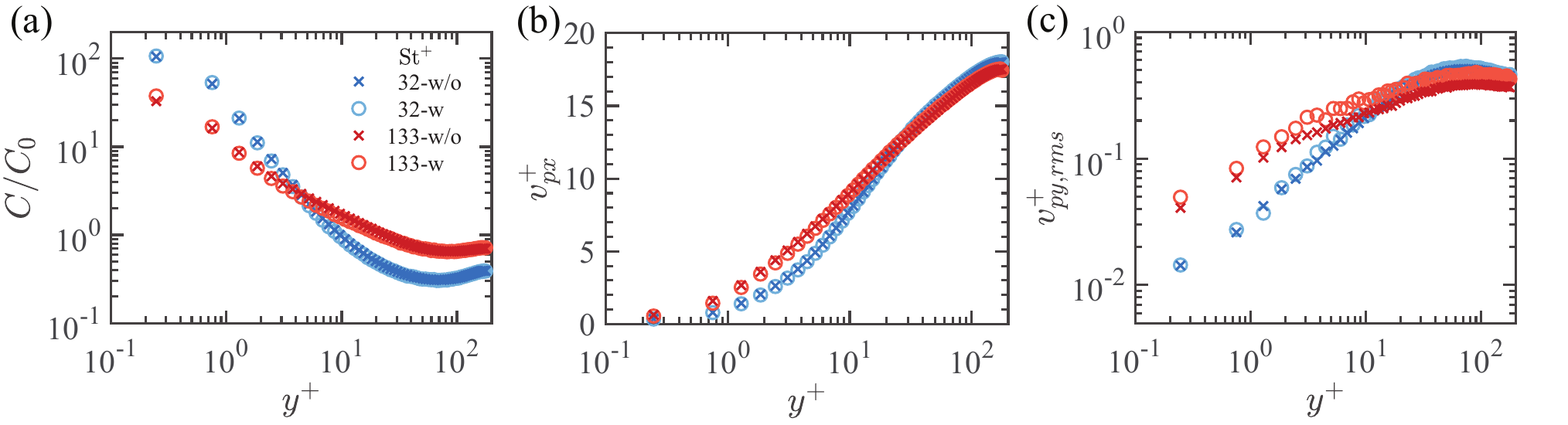}
    \caption{Comparison of (a) normalized wall-normal particle concentration $C/C_0$, (b) mean streamwise particle velocity $v_{px}^+$, and (c) root-mean-square of wall-normal particle velocity $v_{py,rms}^+$ between cases without (w/o) and with (w) the velocity correction.}
    \label{fig:VelocityCorrection}
\end{figure}

Figure \ref{fig:VelocityCorrection} compares typical particle statistics between cases with and without the velocity correction. The RMS of the wall-normal particle velocity, shown in figure \ref{fig:VelocityCorrection}(c), exhibits a slight increase when the velocity correction scheme is applied. The increase in $v_{py,rms}^+$ for $St^+=133$ is also found to be larger than that for $St^+=32$. These changes are reasonable, as the fluid drag force calculated using the undisturbed fluid velocity is larger, making inertial particles more responsive to background turbulence fluctuations and, therefore, more energetic. Moreover, particles with larger inertia ($St^+=133$) generally experience more significant slip velocities, making their statistics more sensitive to velocity correction compared to those with moderate inertia $St^+=32$. However, both the wall-normal concentration (figure \ref{fig:VelocityCorrection}(a)) and the mean streamwise particle velocity (figure \ref{fig:VelocityCorrection}(b)) show no noticeable differences. We thus conclude that errors in drag force calculation do not significantly affect particle transport under current conditions, so the main conclusions of this work remain valid.

\section{Validation of the electrostatic computation}
\label{sec:Ewald}
To validate the PP electrostatic force, we consider the Coulomb force acting on $N_{\mathrm{p}}=5000$ particles in the 3D periodic box with a side length of $L=2 \pi$. Half of the particles carry a nominal positive charge $q=1$, while the others carry a nominal negative charge $q=-1$. For this charge-neutral system, the exact Coulomb force acting on particle $i$ can be computed by the standard Ewald summation \citep{DesernoJChemPhys1998} as

\begin{equation*}
    \boldsymbol{F}^{\mathrm{E,Ewald}}_i = \boldsymbol{F}^{(r)}_i + \boldsymbol{F}^{(k)}_i + \boldsymbol{F}^{(d)}_i,
\end{equation*}

\noindent
where the contribution from the real space $\boldsymbol{F}^{\mathrm{(r)}}_i$, the Fourier space $\boldsymbol{F}^{\mathrm{(k)}}_i$, and the dipole correction $\boldsymbol{F}^{\mathrm{(d)}}_i$ are given as

\begin{equation*}
        \boldsymbol{F}^{\mathrm{(r)}}_i = \frac{q_i}{4 \pi \varepsilon_0} \sum_j q_j \sum_{\boldsymbol{m} \in \mathbb{Z}^3}^{\prime} \left( \frac{2 \alpha}{\sqrt{\pi}} \mathrm{exp} \left( -\alpha^2 |\boldsymbol{r}_{ij} + \boldsymbol{m}L|^2 \right) + \frac{\mathrm{erfc} \left( \alpha |\boldsymbol{r}_{ij} +\boldsymbol{m}L| \right)}{|\boldsymbol{r}_{ij} +\boldsymbol{m}L|} \right) \frac{\boldsymbol{r}_{ij} +\boldsymbol{m}L}{|\boldsymbol{r}_{ij} +\boldsymbol{m}L|^2},
    \end{equation*}

    \begin{equation*}
        \boldsymbol{F}^{\mathrm{(k)}}_i = \frac{q_i}{4 \pi \varepsilon_0 L^3} \sum_j q_j \sum_{\boldsymbol{k} \neq \boldsymbol{0}} \frac{4 \pi \boldsymbol{k}}{k^2} \mathrm{exp} \left( -\frac{k^2}{4 \alpha^2} \right) \sin{(\boldsymbol{k} \cdot \boldsymbol{r}_{ij})},
    \end{equation*}

    \begin{equation*}
        \boldsymbol{F}^{\mathrm{(d)}}_i = -\frac{q_i}{\varepsilon_0 (1+2 \varepsilon^{\prime})L^3} \sum_j q_j \boldsymbol{x}_j.
    \end{equation*}

\noindent
Here, $\alpha$ is the Ewald parameter, $\mathrm{erfc}$ is the complementary error function, and $\varepsilon^{\prime}=1$ is the relative dielectric constant of the surrounding medium. 

\begin{table}
   \begin{center}
\def~{\hphantom{0}}
  \begin{tabular}{ll}
    Parameters & Values\\
    Domain size, $L$ & $2 \pi$\\
    Particle number, $N_{\mathrm{p}}$ & $5000$\\
    Particle charge, $q$ & $\pm 1$\\
    Accuracy parameter, $\alpha r_{\mathrm{C}}$ & $\pi$\\
    Cut-off distance in real space, $r_{\mathrm{C}}$ & $\pi$ \\
    Cut-off wavenumber in Fourier space, $k_{\mathrm{C}}$ & $6$\\
    Error in real space, $\varepsilon^{\mathrm{(r)}}$ & $1.65 \times 10^{-5}$\\
    Error in Fourier space, $\varepsilon^{\mathrm{(k)}}$ & $2.06 \times 10^{-5}$\\
    \end{tabular}
  \caption{Dimensionless parameters for Ewald summation.}
  \label{tab:EwaldSummation}
  \end{center}
\end{table}

Table \ref{tab:EwaldSummation} lists the parameters used in Ewald summation. The dimensionless product $\alpha r_{\mathrm{C}}$ is set to $\pi$ to ensure high accuracy in both real and Fourier spaces. The cut-off radius ($r_{\mathrm{C}}$) and wavenumber ($k_{\mathrm{C}}$) in real and Fourier spaces, respectively, are then determined by $r_{\mathrm{C}}=(\alpha r_{\mathrm{C}}) L/\pi^{1/2}N_{\mathrm{p}}^{1/6}$ and $k_{\mathrm{C}} = 1.8 (\alpha r_{\mathrm{C}})^2/r_{\mathrm{C}}$ to balance the computation cost of $\boldsymbol{F}^{\mathrm{(r)}}_i$ and $\boldsymbol{F}^{\mathrm{(k)}}_i$ \citep{FinchamMS1994}. Due to the high accuracy of Ewald summation, $\boldsymbol{F}^{\mathrm{E,Ewald}}_i$ is used as the reference electrostatic force acting on the particles. In section \ref{sec:P3M}, the P$^3$M method is also employed to compute the electrostatic force under the same conditions. The relative error, $\epsilon_r$, is then evaluated using (\ref{eq:ElectricError}). With the minimum $\epsilon_r<1\%$ (figure \ref{fig:EsMethod}(d)), the P$^3$M method is considered accurate for computing the PP electrostatic force.\\

\section{Derivation of the wall-normal particle concentration profile}
\label{sec:Appd_Model_Derivation}

In this appendix, the wall-normal particle concentration profile is derived following \citet{JohnsonJFM2020}. Define $f(y,v_{py};t)$ as the probability density function of particles in the phase space $(y,v_{py})$ at time $t$, where $y$ is the wall-normal particle location and $v_{py}$ is the wall-normal particle velocity. By definition, the wall-normal particle concentration profile $C(y;t)$ can be directly determined from $f(y,v_{py};t)$ as 

\begin{equation}
    C(y;t)=C_0 \int_{-\infty}^{\infty} f(y,v_{py};t) \mathrm{d} v_{py},
\end{equation}

\noindent
where $C_0$ is the domain-averaged particle concentration. The governing equation of $f$ in the phase space is

\begin{equation}
\label{eq:PhaseSpaceGoverEq}
    \frac{\partial f}{\partial t} + \frac{\partial \left( v_{py} \cdot f \right)}{\partial y} + \frac{\partial \left(a_{py} \cdot f\right)}{\partial v_{py}} = \Dot{f}_{C}.
\end{equation}

\noindent
Here $a_{py}=\mathrm{d} v_{py}/\mathrm{d} t$ is the wall-normal particle acceleration. $\Dot{f}_{C}$ is the change of $f$ due to collisions. Two simplifications can be made here: (i) as particles are assumed to be elastic in this study, both particle mass and momentum are conserved in each collision, which leads to $\Dot{f}_{C}=0$; (ii) when the particle field reaches equilibrium, $f$ is time-independent, i.e., ${\partial f}/{\partial t}=0$. By multiplying the simplified (\ref{eq:PhaseSpaceGoverEq}) by $v_{py}$ and $C_0$, and then integrating from $v_{py}=-\infty$ to $v_{py}=\infty$, the momentum conservation equation can be written as

\begin{equation}
\label{eq:MomentomConservation}
    \frac{\mathrm{d}}{\mathrm{d} y} \left( \langle v_{py}^2 | y \rangle C \right) - \langle a_{py} | y \rangle C = 0.
\end{equation}

\noindent
Here, the notation $\langle *|y \rangle$ denotes the ensemble average of quantities conditioned at the wall-normal location $y$. For a particle located at $y$, its wall-normal acceleration due to drag and electrostatic force is

\begin{equation}
\label{eq:ParticleAcceleration1D}
    a_{py}=\frac{f_I \left(u_{fy}-v_{py} \right)}{\tau_p}+\frac{qE_y}{m}, 
\end{equation}

\noindent
where $u_{fy}$ and $E_y$ are the wall-normal components of the fluid velocity and the electric field. Plugging (\ref{eq:ParticleAcceleration1D}) into (\ref{eq:MomentomConservation}) and integrating along the wall-normal direction then yields the wall-normal particle concentration profile

\begin{equation}
\label{eq:AppdConcentrationProfile}
\begin{split}
    C(y)= \mathcal{C}^{\prime} \mathrm{exp} \Bigg( &  -\int_0^y \frac{\mathrm{d} \ln{ \langle v_{py}^2 | \eta \rangle}}{\mathrm{d} \eta} \mathrm{d} \eta  + \frac{1}{\tau_p} \int_0^y \frac{\langle f_I (u_{fy}-v_{py}) |\eta \rangle}{\langle v_{py}^2 |\eta \rangle} \mathrm{d} \eta  \\
    & + \frac{q}{m} \int_0^y \frac{\langle E_{y} | \eta \rangle}{\langle v_{py}^2 | \eta \rangle} \mathrm{d} \eta \Bigg),
\end{split}
\end{equation}

\noindent
where $\mathcal{C}^{\prime}$ is an unknown coefficient that can be determined from particle mass conservation.

\bibliographystyle{jfm}
\bibliography{jfm-instructions}

\end{document}